\documentclass[12pt]{article}

\usepackage{amsmath,amssymb}           
\usepackage{epsfig}                    
\usepackage[matrix,arrow]{xy}          
\usepackage{xspace}                    
\usepackage{stmaryrd}                  
\usepackage{slashed}

\usepackage{jheppub}                   

\newcommand{\bq}{\begin{equation}}
\newcommand{\eq}{\end{equation}}
\newcommand{\bea}{\begin{eqnarray}}
\newcommand{\eea}{\end{eqnarray}}

\newcommand{\dd}{\mathrm{d}}
\newcommand{\ee}{\mathrm{e}}
\newcommand{\ii}{\mathrm{i}}

\newcommand{\der}{\partial}

\newcommand{\bbZ}{\mathbb{Z}}
\newcommand{\bbR}{\mathbb{R}}
\newcommand{\bbC}{\mathbb{C}}

\DeclareMathOperator{\SU}{\mathit{SU}}
\DeclareMathOperator{\SO}{\mathit{SO}}
\DeclareMathOperator{\SL}{\mathit{SL}}
\DeclareMathOperator{\GL}{\mathit{GL}}

\DeclareMathOperator{\Spin}{\mathit{Spin}}

\DeclareMathOperator{\Cliff}{Cliff}

\newcommand{\id}{1}

\DeclareMathOperator{\tr}{tr}

\newcommand{\Gs}[1]{\Gamma(#1)}

\newcommand{\Lgen}{L}

\newcommand{\Bgen}[2]{\left\llbracket#1,#2\right\rrbracket}
\newcommand{\BLie}[2]{\left[#1,#2\right]}

\newcommand{\Dgen}{{D}}

\DeclareMathOperator{\adj}{ad}

\newcommand{\GM}[2]{\big<#1,#2\big>}

\newcommand{\bisp}[1]{#1_{\!\scriptscriptstyle\#}} 

\newcommand{\LC}{\nabla}

\newcommand{\Riem}{\mathcal{R}}
\newcommand{\Ric}{\mathcal{R}}
\newcommand{\Scalar}{\mathcal{R}}

\newcommand{\GenR}{R}
\newcommand{\GenRic}{R}
\newcommand{\GenS}{S}

\newcommand{\TA}{T_1}
\newcommand{\TV}{T_2}

\DeclareMathOperator{\Diff}{Diff}
\newcommand{\Gsym}{G_{\text{NS}}}

\newcommand{\Ggeom}{G_{\textrm{split}}}

\newcommand{\mukai}[2]{\big<{#1},{#2}\big>}

\newcommand{\Leo}{\Lambda^\text{even/odd}}

\title{Supergravity as Generalised Geometry I: \\ Type II Theories} 

\author{Andr\'e Coimbra\footnote{a.coimbra08@imperial.ac.uk}, 
  Charles Strickland-Constable\footnote{charles.strickland-constable08@imperial.ac.uk}  
  and Daniel Waldram\footnote{d.waldram@imperial.ac.uk}}
\affiliation{Department of Physics,
   Imperial College London \\
   Prince Consort Road, London, SW7 2AZ, UK}

\abstract{
We reformulate ten-dimensional type II supergravity as a generalised
geometrical analogue of Einstein gravity, defined by an
$O(9,1)\times O(1,9)\subset O(10,10)\times\bbR^+$ structure on the
generalised tangent space. Using the notion of generalised connection
and torsion, we introduce the analogue of the Levi--Civita connection,
and derive the corresponding tensorial measures of generalised
curvature. We show how, to leading order in the fermion fields, these
structures allow one to rewrite the action, equations of motion and
supersymmetry variations in a simple, manifestly
$\Spin(9,1)\times\Spin(1,9)$-covariant form. The same formalism also
describes $d$-dimensional compactifications to flat space. 
}

\begin{document}
\maketitle


\section{Introduction}
\label{sec:intro}


Generalised geometry~\cite{GCY,Gualtieri} is the study of structures
on a generalised tangent space $E\simeq TM\oplus T^*M$. Local
diffeomorphism invariance is replaced by a larger group that also
includes the gauge transformations of the NSNS two-form $B$ and there
is a natural $O(d,d)$ structure on $E$, forming a Courant
algebroid~\cite{roytenberg}. Since it was first applied to
supersymmetric type II
backgrounds~\cite{GMPT,JW,Berglund:2005dm} and string sigma
models~\cite{sigmas}, it has been clear that it is closely connected
to the geometry of supergravity. 

In this paper we show that ten-dimensional type IIA and IIB
supergravity theories, to leading order in the fermions, can be
formulated precisely as generalised geometrical analogues of Einstein
gravity. The theory has manifest local $\Spin(9,1)\times\Spin(1,9)$
symmetry and admits a natural analogue of the Levi--Civita
connection. Remarkably both bosonic and fermionic equations of motion
and all the supersymmetry variations take a simple form in terms of
this generalised connection. Although we will focus on the
ten-dimensional case, the same formalism, with a
local $\Spin(d)\times\Spin(d)$ symmetry, also describes 
the type II fields restricted to a $d$-dimensional manifold $M$ used
to compactify the ten-dimensional theory to flat space.  

Interestingly, such rewritings in terms of generalised geometry appear
not to be restricted to type II theories. In a forthcoming companion
paper~\cite{CSW2}, we define the corresponding structures in the
$E_{d(d)}$ version of generalised  geometry~\cite{chris,PW} relevant
to restrictions of eleven-dimensional supergravity to a
$d$-dimensional manifold.  

Let us start by briefly summarizing our construction and results. We
slightly extend the action on the generalised tangent space to a
conformal $O(10,10)\times\bbR^+$ structure. The NSNS fields then
define an $O(9,1)\times O(1,9)$ substructure. The RR field strengths
$F$ are described by a $\Spin(10,10)$ spinor, while the fermions
transform in particular spinor representations of the two $\Spin(9,1)$
groups. The supergravity is described as an analogue of conventional 
gravity, where $O(10,10)\times\bbR^+$ and $O(9,1)\times O(1,9)$ play
the role of the $\GL(d,\bbR)$ and $O(d)$ actions of the frame bundle
respectively, and the diffeomorphism group is replaced by $\Gsym$, an
extension by NSNS $B$-field transformations.  

The central object in the construction is the analogue of the
Levi--Civita connection, a generalised connection $\Dgen$ that is both
compatible with the $O(9,1)\times O(1,9)$ structure and torsion-free
in a generalised sense. Interestingly this connection is not
unique. However, using the $O(9,1)\times O(1,9)$ structure one can
contract indices to construct unique expressions. Using $\Dgen$, the
dynamics and symmetries of the supergravity theories then can be
written in a simple $\Spin(9,1)\times\Spin(1,9)$ covariant form.  
For example, the supersymmetry variations of the gravitini and
dilatini can be written as 
\begin{equation}
\label{eq:intro:susy}
\begin{aligned}
   \delta \psi^+_{\bar{a}} &= 
     \Dgen_{\bar{a}}\epsilon^+ 
     + \tfrac{1}{16} \bisp{F} \gamma_{\bar{a}} \epsilon^- , &&&
   \delta\rho^+ 
      &= \gamma^a\Dgen_a\epsilon^+ , \\
   \delta \psi^-_a &= \Dgen_a\epsilon^- 
   	+ \tfrac{1}{16} \bisp{F}^T  \gamma_{a} \epsilon^+ , &&&
   \delta\rho^- 
      &= \gamma^{\bar{a}}\Dgen_{\bar{a}}\epsilon^- . 
\end{aligned}
\end{equation}
Here $\pm$ and $a$ and $\bar{a}$ refer to spinors and vector indices
respectively of the two $\Spin(9,1)$ groups while $\bisp{F}$ denotes
the RR fields viewed as a $\Spin(9,1)\times\Spin(1,9)$
bispinor. Similarly the bosonic equations of motion can be written as 
\begin{equation}
\label{eq:intro-eom1}
   \GenRic_{a\bar{b}} + \tfrac{1}{16}
	\Phi^{-1}\mukai{F}{\Gamma_{a \bar{b}} F} = 0 , \qquad
   \GenS = 0 , \qquad
   \Gamma^A\Dgen_A F=0 , 
\end{equation}
where $\GenRic$ and $\GenS$ are generalised curvature tensors
constructed from $\Dgen$ and $\Gamma^A$ are the $\Spin(10,10)$ gamma
matrices. (The structure of the quadratic RR field term and $\Phi$ are
explained in the main text.) The  fermionic equations of motion are 
\begin{equation}
\label{eq:intro-eom2}
\begin{aligned}
   \gamma^b \Dgen_b \psi^+_{\bar{a}} - \Dgen_{\bar{a}} \rho^+ 
       &= \tfrac{1}{16} \gamma^b \bisp{F} 
          \gamma_{\bar{a}} \psi^-_b  , &&&
   \gamma^a \Dgen_a \rho^+ - \Dgen^{\bar{a}} \psi^+_{\bar{a}}
       &= - \tfrac{1}{16} \bisp{F} \rho^-  ,\\
   \gamma^{\bar{b}} \Dgen_{\bar{b}} \psi^-_{a} - \Dgen_{a} \rho^- 
       &= \tfrac{1}{16}  \gamma^{\bar{b}} \bisp{F}^T 
           \gamma_{a} \psi^+_{\bar{b}}  , &&&
   \gamma^{\bar{a}} \Dgen_{\bar{a}} \rho^- - \Dgen^a \psi^-_a
       &= - \tfrac{1}{16} \bisp{F}^T \rho^+ .
\end{aligned}
\end{equation}
and there are similar covariant expressions for the bosonic
supersymmetry variations and the action. 

The idea that supergravity can be reformulated with larger local
symmetry groups and with a structure reflecting the duality symmetries
of string theory is not new and there are several precursors of
the work reported here and significant related formulations. In the mid-80s,
de Wit and Nicolai considered larger structures and local symmetries
related to U-duality groups in the context of eleven-dimensional
supergravity in~\cite{deWN,Nic,KNS} and the generalised geometry discussed
here can be viewed as the geometrical basis for their formalism in the
type II context. (In~\cite{CSW2} we will directly address the case of
local $SU(8)$ symmetry considered in their original work.) Focusing on
the NSNS sector, 
in~\cite{Siegel:1993xq,Siegel:1993th} Siegel introduced a doubled
$2d$-dimensional tangent space (on a doubled, though restricted,
spacetime) with a local $\GL(d,\bbR)\times\GL(d,\bbR)$ symmetry. He
further introduced connections and curvatures, essentially defining
the $\GenRic$ and $\GenS$ curvatures that appear
in~\eqref{eq:intro-eom1}. 

More recently, Hull and Zwiebach introduced
``double field theory''~\cite{Hull:2009mi}, motivated by
$2d$-dimensional target space models of non-geometrical
backgrounds~\cite{Tfold}. In this and subsequent
work~\cite{Hull:2009zb,Hohm:2010jy,Hohm:2010pp,Kwak:2010ew}, the NSNS
action is formulated in terms of first-order derivatives of doubled
objects and from this equations of motion are obtained. The relation
to Siegel's formalism was made in~\cite{Hohm:2010pp} and then expanded on
in~\cite{Hohm:2010xe}. A closely related construction in terms of curvatures
of ``semi-covariant'' derivatives on the doubled space was given
in~\cite{JLP}. The result that the RR action can be rewritten in terms
of $\Spin(d,d)$ spinors goes back to~\cite{BMZ,Fukuma:1999jt}. This
and the $\Spin(d,d)$ formulation of the RR equations of motion were  
both considered in the double field theory formalism by Hull, who, in
addition, studied the $\Spin(9,1)\times\Spin(1,9)$ field 
representations used in this paper from the doubled world-sheet
perspective~\cite{hull1}\footnote{While completing this paper, we also 
  received~\cite{Hohm:2011zr} which extends the double field theory
  formalism to include the RR fields and has some overlap with the
  results here.}. 

In earlier work, West conjectured that a non-linear realisation of
$E_{11}$ underlies M theory~\cite{West}. In the context of type 
II supergravity~\cite{WestII}, the same first-order NSNS action was
recently derived in~\cite{West:2010ev} and then extended to describe
the RR equations of motion in~\cite{Rocen:2010bk}. Note that similar
actions relevant to restrictions of eleven-dimensional supergravity
to lower-dimensional manifolds are also discussed
in~\cite{Berman:2010is} and there are further related works
in~\cite{Hohm:2011dz,Copland:2011yh,ALLP,thompson}. 

Siegel's formalism and that of double field theory start with fields
depending on $2d$ coordinates but then impose a constraint that means,
on a given coordinate patch, all the fields are independent of half
the coordinates (as does West's approach). Thus locally the starting
point for these formalisms is the same as that of generalised
geometry. In particular, the definition of the generalised connection 
$\Dgen$ given here can be directly applied to the double field theory
formalism. We will comment briefly on these relationships in the
conclusions.  

In the mathematics literature, the basic notion of the generalised
tangent space with an $O(d,d)$ metric and a suitable bracket is known
as an exact Courant algebroid (see~\cite{roytenberg,CSX} and
references therein). Additional ``generalised geometry'' structures on
such objects, specifically generalised complex structures and $O(d)\times
O(d)$ generalised metrics, were introduced by Hitchin and 
Gualtieri~\cite{GCY,Gualtieri}. Connections on Courant algebroids were
introduced in~\cite{AX} (see also~\cite{CSX}) and again
in~\cite{ellwood} and~\cite{gualtieri-conn}, together with a notion of
torsion and compatibility with the generalised metric. 

The paper is arranged as follows. Section~\ref{sec:NSNS} summarises type IIA and
IIB supergravity in the democratic formalism~\cite{democratic}, using
slightly modified notation, a new linear combination of dilatini and
gravitini fields, and a significant rewriting of the fermionic
terms. We also discuss the patching of the NSNS $B$-field and the
symmetry algebra of the NSNS sector, both of which are reflected in
the generalised geometry. Sections~\ref{sec:OddGG} and~\ref{sec:OdOd-sec} introduce the key
concepts of generalised geometry that we will use and show that one
can always construct a torsion-free, $O(p,q)\times O(q,p)$-compatible
generalised connection $\Dgen$, the analogue of the Levi--Civita
connection in Riemannian geometry. In section~\ref{sec:IIasgen} we then rewrite the
type II action, supersymmetry variations and equations of motion using
these new geometrical constructions. We conclude with some summary and
discussion in section~\ref{sec:conc}.


\section{Type II supergravity}
\label{sec:NSNS}

Let us briefly recall the structure of $d=10$ type II supergravity. We
essentially follow the conventions of the democratic
formalism~\cite{democratic}, as summarised in appendix~\ref{app:conv},
and consider only the leading-order fermionic terms. We introduce a
slightly unconventional notation in a few places in order to match
more naturally with the underlying generalised geometry. It is also
helpful to considerably rewrite the fermionic sector, introducing a
particular linear combination of dilatini and gravitini, to match more
closely what follows. 


\subsection{Degrees of freedom, equations of motion and supersymmetry}
\label{sec:eom}

The type II fields are denoted 
\begin{equation}
   \{ g_{\mu\nu}, B_{\mu\nu},\phi, A^{(n)}_{\mu_1\dots \mu_n}, 
       \psi_\mu^\pm, \lambda^\pm \} , 
\end{equation}
where $g_{\mu\nu}$ is the metric, $B_{\mu\nu}$ the two-form potential, $\phi$
is the dilaton and $A^{(n)}_{\mu_1\dots \mu_n}$ are the RR potentials 
in the democratic formalism, with $n$ odd for type IIA and $n$ even
for type IIB.  In each theory there is also a pair of
chiral gravitini $\psi_\mu^\pm$ and a pair of chiral dilatini
$\lambda^\pm$. Here our notation is that $\pm$ does \emph{not} refer
to the chirality of the spinor but, as we will see, denote generalised
geometrical subspaces. Specifically, in the notation
of~\cite{democratic}, for type IIA they are the chiral components of
the gravitino and dilatino 
\begin{equation}
\begin{aligned}
\label{eq:IIAfermi}
   \psi_\mu &= \psi^+_\mu + \psi^-_\mu && &&
        \text{where} & \gamma^{(10)}\psi^\pm_\mu &=\mp\psi^\pm_\mu  \\
   \lambda &= \lambda^+ + \lambda^- && &&
        \text{where} & \gamma^{(10)}\lambda^\pm &= \pm\lambda^\pm . 
\end{aligned}
\end{equation}
(Note that $\psi_\mu^+$ and $\lambda^+$, and similarly $\psi_\mu^-$ and
$\lambda^-$, have \emph{opposite} chiralities.) For type IIB, in the
notation of~\cite{democratic} one has two component objects 
\begin{equation}
\begin{aligned}
\label{eq:IIBfermi}
   \psi_\mu 
      &= \begin{pmatrix} \psi_\mu^+ \\ \psi_\mu^- \end{pmatrix} && &&
        \text{where} & \gamma^{(10)}\psi^\pm_\mu &= \psi^\pm_\mu  \\
   \lambda 
      &= \begin{pmatrix} \lambda^+ \\ \lambda^- \end{pmatrix} && &&   
        \text{where} & \gamma^{(10)}\lambda^\pm &= - \lambda^\pm . 
\end{aligned}
\end{equation}
and again the gravitini and dilatini have opposite chiralities. 

In what follows, it will be very useful to consider the quantities 
\begin{equation}
\label{eq:rho-def}
   \rho^\pm := \gamma^\mu \psi_\mu^\pm - \lambda^\pm ,  
\end{equation}
instead of $\lambda^\pm$. These are the natural combinations that
appear in generalised geometry and from now on we will use $\rho^\pm$
rather than $\lambda^\pm$. 

The bosonic ``pseudo-action'' takes the form
\begin{equation}
\label{eq:NSaction}
   S_{\text{B}} = \frac{1}{2\kappa^2}\int \sqrt{-g}\,\Big[
         \ee^{-2\phi} \left( 
            \Scalar + 4 (\der\phi)^2 - \tfrac{1}{12}H^2 \right) 
          - \tfrac{1}{4}\sum_n \tfrac{1}{n!}(F^{(B)}_{(n)})^2 \Big], 
\end{equation}
where $H=\dd B$ and $F^{(B)}_{(n)}$ is the $n$-form RR field strength. Here
we will use the ``$A$-basis'', where the field strengths, as sums of even or odd forms, take the form\footnote{Note that
  in type IIA one cannot write a potential for the zero-form field
  strength, which must instead be added by hand in~\eqref{eq:RR}. Note
  also that in~\cite{democratic} these field strengths are denoted $G$.}  
\begin{equation}
\label{eq:RR}
   F^{(B)} = \sum_n F^{(B)}_{(n)} = \sum_n\ee^B \wedge \dd A_{(n-1)} ,
\end{equation}
where $\ee^B=1+B+\frac{1}{2}B\wedge B+\dots$. This is a
``pseudo-action'' because the RR fields satisfy a self-duality relation
that does not follow from varying the action, namely,
\begin{equation}
   \label{eq:RR-duality}
   F^{(B)}_{(n)} = (-)^{[n/2]} * F^{(B)}_{(10-n)} , 
\end{equation}
where $[n]$ denotes the integer part and ${}*\omega$ denotes the Hodge
dual of $\omega$. The fermionic action, keeping only terms quadratic
in the fermions, can be written after some manipulation as
\begin{equation}
\label{eq:fermi-action}
\begin{aligned}
   S_{\text{F}} &= -\frac{1}{2\kappa^2}\int \sqrt{-g} \Big[\ee^{-2\phi} 
      \Big( 
	2 \bar\psi^{+\mu} \gamma^\nu \LC_\nu \psi^+_\mu 
           - 4 \bar\psi^{+\mu} \LC_\mu \rho^+
           - 2 \bar\rho^+ \slashed{\LC} \rho^+
      \\ & \qquad \qquad \qquad 
           - \tfrac12 \bar\psi^{+\mu} \slashed{H} \psi^+_\mu 
           - \bar\psi^+_\mu H^{\mu\nu\lambda} \gamma_\nu \psi^+_\lambda
           - \tfrac12 \rho^+ H^{\mu\nu\lambda}\gamma_{\mu\nu}
              \psi^+_\lambda 
           + \tfrac12 \rho^+ \slashed{H} \rho^+ \Big)
      \\ & \qquad \qquad + \ee^{-2\phi} \Big(
        2 \bar\psi^{-\mu} \gamma^\nu \LC_\nu \psi^-_\mu 
           - 4 \bar\psi^{-\mu} \LC_\mu \rho^-
           - 2 \bar\rho^- \slashed{\LC} \rho^- 
      \\ & \qquad \qquad \qquad
           + \tfrac12 \bar\psi^{-\mu} \slashed{H} \psi^-_\mu 
           + \bar\psi^-_\mu H^{\mu \nu \lambda} \gamma_\nu \psi^-_\lambda
           + \tfrac12 \rho^-H^{\mu\nu\lambda}\gamma_{\mu\nu} 
              \psi^-_\lambda 
           - \tfrac12 \rho^- \slashed{H} \rho^- \Big)
      \\ & \qquad \qquad
	- \tfrac14 \ee^{-\phi} \Big( 
             \bar\psi^+_\mu\gamma^\nu \slashed{F}^{(B)}
                 \gamma^\mu\psi^-_\nu
             + \rho^+ \slashed{F}^{(B)} \rho^- 
             \Big)
        \Big] . 
\end{aligned}
\end{equation}
where $\LC$ is the Levi--Civita connection. 

To match what follows it is useful to rewrite the standard equations
of motion in a particular form. For the bosonic fields, with the
fermions set to zero, one takes the combinations that naturally arise
from the string $\beta$-functions, namely 
\begin{equation}
\label{eq:bose-eom}
\begin{aligned} 
   \Ric_{\mu\nu} - \tfrac{1}{4}H_{\mu\lambda\rho}H_\nu{}^{\lambda\rho} 
      + 2 \nabla_\mu\nabla_\nu \phi 
      - \tfrac{1}{4}\ee^{2\phi} \sum_n \tfrac{1}{(n-1)!} 
         F^{(B)}_{\mu \lambda_1\dots \lambda_{n-1}}
         F_{\, \nu}^{(B)\lambda_1\dots \lambda_{n-1}} 
      &= 0 ,  \\
   \LC^\mu\left(\ee^{-2\phi}H_{\mu \nu \lambda}\right) 
      - \tfrac{1}{2}\sum_n \tfrac{1}{(n-2)!} 
         F^{(B)}_{\mu \nu \lambda_1\dots \lambda_{n-2}}
         F^{(B)\lambda_1\dots \lambda_{n-2}} 
      &= 0 ,  \\
   \LC^2\phi - \left(\LC\phi\right)^2
      + \tfrac{1}{4}\Scalar - \tfrac{1}{48} H^2 &= 0 ,  \\
   \dd F^{(B)} - H\wedge F^{(B)} &= 0 , 
\end{aligned}   
\end{equation}
where the final Bianchi identity for $F$ follows from the
definition~\eqref{eq:RR}. Keeping only terms linear in the
fermions, the fermionic equations of motion read
\begin{equation}
\label{eq:fermi-eom}
\begin{aligned}
   & \gamma^\nu\left[ \left(\LC_\nu 
	\mp \tfrac{1}{24} H_{\nu\lambda\rho} \gamma^{\lambda\rho} 
	- \der_\nu \phi \right) \psi^\pm_\mu 
        \pm \tfrac{1}{2} H_{\nu\mu}{}^\lambda \psi^\pm_\lambda \right]
        - \left( \LC_\mu \mp \tfrac{1}{8} H_{\mu\nu\lambda} \gamma^{\nu\lambda}
        \right)\rho^\pm 
        \\ & \qquad \qquad \qquad 
        = \tfrac{1}{16}\ee^{\phi}\sum_n (\pm)^{[(n+1)/2]}
            \gamma^\nu \slashed{F}^{(B)}_{(n)} \gamma_\mu \psi^\mp_\nu , \\
   & \left(\LC_\mu \mp \tfrac{1}{8} H_{\mu\nu\lambda} \gamma^{\nu\lambda}
           - 2 \der_\mu \phi \right) \psi^{\mu \pm}
        - \gamma^\mu \left( \LC_\mu 
           \mp \tfrac{1}{24}H_{\mu\nu\lambda} \gamma^{\nu\lambda}
           - \der_\mu \phi \right) \rho^\pm
        \\ & \qquad \qquad \qquad 
        = \tfrac{1}{16}\ee^{\phi}\sum_n (\pm)^{[(n+1)/2]}
           \slashed{F}^{(B)}_{(n)} \rho^\mp , 
\end{aligned}
\end{equation}

The supersymmetry variations are parametrised by are pair of chiral
spinors $\epsilon^\pm$ where, again, in the notation of~\cite{democratic},
for type IIA, we have 
\begin{equation}
\begin{aligned}
      \epsilon &= \epsilon^+ + \epsilon^- && &&
        \text{where} & \gamma^{(10)}\epsilon^\pm &= \mp\epsilon^\pm ,
\end{aligned}
\end{equation}
while for type IIB we have the doublet
\begin{equation}
\begin{aligned}
   \epsilon 
      &= \begin{pmatrix} \epsilon^+ \\ \epsilon^- \end{pmatrix} && &&   
        \text{where} & \gamma^{(10)}\epsilon^\pm &= \epsilon^\pm . 
\end{aligned}
\end{equation}
Again keeping only linear terms in the fermions field, the
supersymmetry transformations for the bosons read 
\begin{equation}
\label{eq:susy-bos}
\begin{aligned}
   \delta e_\mu^a 
      &= \bar{\epsilon}^+ \gamma^a \psi^+_\mu 
           + \bar{\epsilon}^- \gamma^a \psi^-_\mu , \\ 
   \delta B_{\mu\nu} 
      &=  2\bar{\epsilon}^+ \gamma^{\phantom{+}}_{[\mu} \psi^+_{\nu]} 
           - 2\bar{\epsilon}^- \gamma^{\phantom{+}}_{[\mu} \psi^-_{\nu]} , \\ 
   \delta\phi - \tfrac{1}{4}\delta\log(-g)
      &= - \tfrac{1}{2}\bar{\epsilon}^+ \rho^+
           - \tfrac{1}{2}\bar{\epsilon}^- \rho^- , \\
   \left(\ee^B\wedge \delta A\right)^{(n)}_{\mu_1\dots \mu_n} 
      &= \tfrac{1}{2} \left( \ee^{-\phi} \bar{\psi}^+_\nu
             \gamma_{\mu_1\dots \mu_n} \gamma^\nu \epsilon^-
           - \ee^{-\phi} \bar{\epsilon}^+ \gamma_{\mu_1\dots \mu_n} \rho^- \right)
             \\ & \qquad
          \mp \tfrac{1}{2} \left( \ee^{-\phi}\bar{\epsilon}^+
             \gamma^\nu\gamma_{\mu_1\dots \mu_n}\psi^-_\nu 
          + \ee^{-\phi}\bar{\rho}^+ \gamma_{\mu_1\dots \mu_n} \epsilon^- \right) ,
\end{aligned}
\end{equation}
where $e_\mu$ is an orthonormal frame for $g_{\mu\nu}$ and in the last
equation the upper sign refers to type IIA and the lower to type
IIB. For the fermions one has 
\begin{equation}
\label{eq:susy-fer}
\begin{aligned}
   \delta \psi^\pm_\mu &=
      \left( {\LC}_\mu \mp \tfrac{1}{8}H_{\mu\nu\lambda}\gamma^{\nu\lambda} 
         \right) \epsilon^\pm
         + \tfrac{1}{16}\ee^\phi \sum_n (\pm)^{[(n+1)/2]}
              \slashed{F}^{(B)}_{(n)} \gamma_\mu\epsilon^\mp, \\
   \delta \rho^\pm &=
      \gamma^\mu \left( \LC_\mu 
         \mp \tfrac{1}{24}H_{\mu\nu\lambda}\gamma^{\nu\lambda} - \der_\mu\phi 
         \right) \epsilon^\pm .
\end{aligned}
\end{equation}
%


\subsection{Bosonic symmetries}
\label{sec:symm}

It is useful to recall the symmetries of the NSNS bosonic sector since
these will be reflected in the generalised geometry. The potential $B$
is only locally defined, so that, given an open cover $\{U_i\}$,
across coordinate  patches $U_i\cap U_j$ it can be patched via
\begin{equation}
\label{eq:Bpatch}
   B_{(i)} = B_{(j)} - \dd\Lambda_{(ij)} . 
\end{equation}
Furthermore the one-forms $\Lambda_{(ij)}$ satisfy
\begin{equation}
   \Lambda_{(ij)} 
      + \Lambda_{(jk)}
      + \Lambda_{(ki)} = \dd \Lambda_{(ijk)} ,
\end{equation}
on $U_i\cap U_j\cap U_k$. This makes $B$ a ``connection structure on a
gerbe''~\cite{gerbes}\footnote{In supergravity, there is no
  requirement that the flux $H$ is quantised. However, string theory
  implies the cohomological condition $H/(8\pi^2\alpha')\in
  H^3(M,\bbZ)$ (up to torsion terms). This can be implemented in the
  gerbe structure by requiring
  $g_{ijk}=\exp(4\pi\alpha'\ii\Lambda_{(ijk)})$ satisfy the cocycle
  condition $g_{jkl}g^{-1}_{ikl}g_{ijl}g^{-1}_{ijk}=1$ on $U_i\cap
  U_j\cap U_k\cap U_l$. We will not consider this further restriction
  in the following.}. 
There is a similar patching for the sum of the RR potentials $A$. We
are using the ``$A$-basis'', so, given the field
strengths~\eqref{eq:RR} are globally defined we have, as a sum of
forms \footnote{Note here $i$ and $j$ refer to the patch not the
  degree of the form.},  
\begin{equation}
\label{eq:Apatch}
   A_{(i)} = \ee^{\dd\Lambda_{(ij)}}\wedge A_{(j)} 
      + \dd\hat{\Lambda}_{(ij)} ,
\end{equation}
where $\hat{\Lambda}_{(ij)}$ is a sum of even or odd forms in type
IIA and type IIB respectively.  

Focusing on the NSNS sector symmetry algebra we see that, in addition
to diffeomorphism invariance, we have the local bosonic gauge 
symmetry  
\begin{equation}
\label{eq:Btranfs}
   B'_{(i)} = B_{(i)} - \dd \lambda_{(i)} , \qquad
   A'_{(i)} = \ee^{\dd\lambda_{(i)}}A_{(i)} , 
\end{equation}
where the choice of sign in the gauge transformation is to match the
generalised geometry conventions that follow. Given the patching of
$B$, the only requirement is $\dd\lambda_{(i)}=\dd\lambda_{(j)}$ on
$U_i\cap U_j$. Thus globally $\lambda_{(i)}$ is equivalent to
specifying a closed two-form. The set of gauge symmetries is then the
Abelian group of closed two-forms under addition
$\Omega^2_{\text{cl}}(M)$. The gauge transformations do not commute
with the diffeomorphisms so the NSNS bosonic symmetry group $\Gsym$
has a fibred structure 
\begin{equation}
\label{eq:ff}
   \Omega^2_{\text{cl}}(M) \longrightarrow \Gsym 
      \longrightarrow \Diff(M) , 
\end{equation}
sometimes written as the semi-direct product
$\Diff(M)\ltimes\Omega^2_{\text{cl}}(M)$. 

One can see this structure infinitesimally by combining the
diffeomorphism and gauge symmetries, given a vector $v$ and one-form
$\lambda_{(i)}$, into a general variation 
\begin{equation}
\label{eq:NSsyms}
      \delta_{v+\lambda} g = \mathcal{L}_v g , \qquad
      \delta_{v+\lambda} \phi = \mathcal{L}_v \phi , \qquad
      \delta_{v+\lambda} B_{(i)} 
         = \mathcal{L}_v B_{(i)} - \dd \lambda_{(i)} , 
\end{equation}
where the patching~\eqref{eq:Bpatch} of $B$ implies that 
\begin{equation}
\label{eq:l-patch}
   \dd\lambda_{(i)} = \dd\lambda_{(j)} 
      - \mathcal{L}_v\dd\Lambda_{(ij)} . 
\end{equation}
Recall that $\lambda_{(i)}$ and $\lambda_{(i)}+\dd\phi_{(i)}$ define
the same gauge transformation. One can use this ambiguity to
integrate~\eqref{eq:l-patch} and set 
\begin{equation}
\label{eq:l-patch-new}
   \lambda_{(i)} = \lambda_{(j)} - i_v\dd\Lambda_{(ij)} , 
\end{equation}
on $U_i\cap U_j$. 


\section{$O(d,d)\times\bbR^+$ generalised geometry}
\label{sec:OddGG}

We would like to define the generalised geometric analogues of each
of the ingredients in the construction of the Levi--Civita
connection\footnote{These ingredients are reviewed in
  appendix~\ref{app:LC}.}. In this section we review the
generalisations of the frame bundle, the Lie derivative, connections,
torsion and curvature. In the following section we discuss the notion
of a generalised metric and the analogue of the Levi--Civita connection. 

One way to view generalised geometry is as a formalism for
``geometrising'' the bosonic structures that appear in
supergravity. In the context of the NSNS sector this means first combining
the symmetry algebra of diffeomorphisms and $B$-field gauge
transformations into an algebra of ``generalised'' Lie
derivatives. This structure is known as an  ``exact Courant
algebroid'' in the mathematics
literature~\cite{roytenberg,CSX} and, on a $d$-dimensional manifold,
defines a bundle with a natural $O(d,d)$ action. Combining $g$, $B$
and $\phi$ into a single geometrical object introduces an
additional refinement of the structure, defining a generalised
geometry~\cite{GCY,Gualtieri}. The only slight, though important,
extension we will require here is to promote the $O(d,d)$ action to
$O(d,d)\times\bbR^+$~\cite{GMPW,Jeschek:2005ek}. 


\subsection{Generalised structure bundle}
\label{sec:gen-struc}

We start by recalling the generalised tangent space and defining what
we will call the ``generalised structure'' which is the analogue of
the frame bundle $F$ in conventional geometry. 

Let $M$ be a $d$-dimensional spin manifold. In line with the
patching of the transformation parameters~\eqref{eq:l-patch-new}, one
starts by defining the generalised tangent space $E$. It is defined as
an extension of the tangent space by the cotangent space 
\begin{equation}
\label{eq:Odd-Edef}
   0 \longrightarrow T^*M \longrightarrow E 
      \longrightarrow TM \longrightarrow 0 ,
\end{equation}
which depends on the patching one-forms $\Lambda_{(ij)}$. If
$v_{(i)}\in \Gs{TU_i}$ and $\lambda_{(i)}\in \Gs{T^*U_i}$, so
$V_{(i)}=v_{(i)}+\lambda_{(i)}$ is a section of $E$ over the patch 
$U_i$, then 
\begin{equation}
\label{eq:Odd-patch}
   v_{(i)} + \lambda_{(i)} 
      = v_{(j)} + \big( \lambda_{(j)} -
        i_{v_{(j)}}\dd\Lambda_{(ij)}\big)  ,
\end{equation}
on the overlap $U_i\cap U_j$. Hence as defined, while the $v_{(i)}$
globally are equivalent to a choice of vector, the $\lambda_{(i)}$ do
not globally define a one-form. $E$ is in fact isomorphic to $TM\oplus
T^*M$ though there is no canonical isomorphism. Instead one must choose
a splitting of the sequence~\eqref{eq:Odd-Edef} as discussed below.
Crucially the definition of $E$ is consistent with an $O(d,d)$ metric
given by, for $V=v+\lambda$ 
\begin{equation}
\label{eq:Odd-metric}
   \GM{V}{V} = i_v \lambda , 
\end{equation}
since $i_{v_{(i)}}\lambda_{(i)}=i_{v_{(j)}}\lambda_{(j)}$ on $U_i\cap
U_j$. 

In order to describe the dilaton correctly we will actually need to
consider a slight generalisation of $E$. We define the bundle
$\tilde{E}$ weighted by $\det T^*M$ so that  
\begin{equation}
\label{eq:Ep-def}
   \tilde{E} = \det T^*M \otimes E . 
\end{equation}
The point is that, given the metric~\eqref{eq:Odd-metric}, one can now
define a natural principal bundle with fibre $O(d,d)\times\bbR^+$ in
terms of bases of $\tilde{E}$. We define a
\emph{conformal basis} $\{\hat{E}_A\}$ with $A=1,\dots 2d$ on
$\tilde{E}_x$ as one satisfying  
\begin{equation}
\label{eq:conformal}
   \GM{\hat{E}_A}{\hat{E}_B} = \Phi^2 \eta_{AB} \quad
   \text{where} \quad 
   \eta = \frac{1}{2}\begin{pmatrix} 
      0 & \id \\ \id & 0 \end{pmatrix} .  
\end{equation}
That is $\{\hat{E}_A\}$ is orthonormal up to a frame-dependent conformal
factor $\Phi\in\Gs{\det{T^*M}}$.  We then define the \emph{generalised
  structure bundle} 
\begin{equation}
\label{eq:gen-fb}
   \tilde{F} = \big\{ (x,\{\hat{E}_A\}) : \text{$x\in M$, and 
        $\{\hat{E}_A\}$  is a conformal basis of $\tilde{E}_x$} 
      \big\} . 
\end{equation}
By construction, this is a principal bundle with fibre
$O(d,d)\times\bbR^+$. One can make a change of basis 
\begin{equation}
\label{eq:Mdef}
   V^A \mapsto V^{\prime A} = M^A{}_B V^B , \qquad
   \hat{E}_A\mapsto \hat{E}'_A=\hat{E}_B(M^{-1})^B{}_A .
\end{equation}
where $M\in O(d,d)\times\bbR^+$ so that
$(M^{-1})^C{}_A(M^{-1})^D{}_B\eta_{CD}=\sigma^2\eta_{AB}$ for some
$\sigma$. The topology of $\tilde{F}$ encodes both the
topology of the tangent bundle $TM$ and of the $B$-field gerbe. 

Given the definition~\eqref{eq:Odd-Edef} there is one natural
conformal basis defined by the choice of coordinates on $M$, namely
$\{\hat{E}_A\}=\{\der/\der x^\mu\}\cup\{\dd x^\mu\}$. Given
$V\in\Gs{E}$ over the patch $U_i$, we have $V=v^\mu(\der/\der
x^\mu)+\lambda_\mu\dd x^\mu$, we will sometime denote the
components of $V$ in this frame by an index $M$ such that 
\begin{equation}
\label{coord-frame-cpts}
   V^M = \begin{cases} v^\mu & \text{for $M=\mu$} \\
           \lambda_\mu &  \text{for $M=\mu+d$} 
        \end{cases} . 
\end{equation}
%


\subsection{Generalised tensors and spinors and split frames} 
\label{sec:gen-tensor}

Generalised tensors are simply sections of vector bundles
constructed from different representations of $O(d,d)\times\bbR^+$,
that is representations of $O(d,d)$ of definite weight under
$\bbR^+$. Since the $O(d,d)$ metric gives an isomorphism between $E$
and $E^*$, one has the bundle
\begin{equation}
   E^{\otimes n}_{(p)} 
      = (\det{T^*M})^p\otimes E \otimes \dots \otimes E . 
\end{equation}
for a general tensor of weight $p$. 

One can also consider $\Spin(d,d)$ spinor
representations~\cite{Gualtieri}. The $O(d,d)$ Clifford algebra 
\begin{equation}
   \left\{ \Gamma_A, \Gamma_B \right\} = 2 \eta_{AB} . 
\end{equation}
can be realised on each coordinate patch $U_i$ by identifying spinors
with weighted sums of forms $\Psi_{(i)}\in
\Gs{(\det{T^*U_i})^{1/2}\otimes \Lambda^\bullet T^*U_i}$,
with the Clifford action 
\begin{equation}
   V^A\Gamma_A \Psi_{(i)} 
      = i_v \Psi_{(i)} + \lambda_{(i)}\wedge\Psi_{(i)} .  
\end{equation}
The patching~\eqref{eq:Odd-patch} then implies
\begin{equation}
\label{eq:Odd-spinor-patch}
   \Psi_{(i)} = \ee^{\dd\Lambda_{(ij)}} \wedge \Psi_{(j)} . 
\end{equation}
Projecting onto the chiral spinors then defines two $\Spin(d,d)$ spinor
bundles, isomorphic to weighted sums of odd or even forms 
$S^\pm(E)\simeq(\det{T^*M})^{-1/2}\otimes\Leo T^*M$ where
again specifying the isomorphism requires a choice of splitting. 

More generally one defines $\Spin(d,d)\times\bbR^+$ spinors of weight $p$
as sections of 
\begin{equation}
   S^\pm_{(p)} = (\det{T^*M})^p\otimes S^\pm(E) .
\end{equation}
Note that there is a natural $\Spin(d,d)$ invariant bilinear on these spinor
spaces given by the Mukai pairing~\cite{GCY,Gualtieri}. For $\Psi,
\Psi'\in\Gs{S^\pm_{(p)}}$ 
one has 
\begin{equation}
\label{eq:mukai}
   \mukai{\Psi}{\Psi'} 
      = \sum_n (-)^{[(n+1)/2]} \Psi^{(d-n)} \wedge \Psi^{\prime(n)}
      \in \Gs{(\det{T^*M})^{2p}} ,
\end{equation}
where $\Psi^{(n)}$ and $\Psi^{\prime (n)}$ are the local weighted
$n$-form components. 

A special class of conformal frames are those defined by a splitting
of the generalised tangent space $E$. A splitting is a map $TM\to
E$. It is equivalent to specifying a local two-form $B$ patched as
in~\eqref{eq:Bpatch} and defines an isomorphism $E\simeq TM \oplus
T^*M$. If $\{\hat{e}_a\}$ is a generic basis for $TM$ and $\{e^a\}$ be
the dual  basis on $T^*M$, one can then define what we call a
\emph{split frame} $\{\hat{E}_A\}$ for $\tilde{E}$ by  
\begin{equation}
\label{eq:geom-basis}
   \hat{E}_A = \begin{cases} 
           \hat{E}_a = (\det e) \left(\hat{e}_a + i_{\hat{e}_a} B \right) 
               & \text{for $A=a$} \\
           E^a = (\det e) e^a &  \text{for $A=a+d$}
         \end{cases} . 
\end{equation}
We immediately see that 
\begin{equation}
   \GM{\hat{E}_A}{\hat{E}_B} = (\det e)^2 \eta_{AB} ,
\end{equation}
and hence the basis is conformal. Writing
$V=v^a\hat{E}_a+\lambda_a E^a \in\Gs{\tilde{E}}$ we have
\begin{equation}
\begin{aligned}
   V^{(B)} &= v^a (\det{e})\hat{e}_a + \lambda_a (\det{e})e^a \\
     &= v_{(i)} + \lambda_{(i)} - i_{v_{(i)}} B_{(i)} ,
\end{aligned}
\end{equation}
demonstrating that the splitting defines an isomorphism
$\tilde{E}\simeq(\det{T^*M})\otimes(TM \oplus T^*M)$ since
$\lambda_{(i)} - i_{v_{(i)}} B_{(i)}=\lambda_{(j)} - i_{v_{(j)}}
B_{(j)}$. 

The class of split frames defines a sub-bundle of $\tilde{F}$. Such
frames are related by transformations~\eqref{eq:Mdef} where $M$ takes
the form
\begin{equation}
\label{eq:Ggeom}
   M = (\det A)^{-1}
       \begin{pmatrix} \id & 0  \\ \omega & \id \end{pmatrix}
       \begin{pmatrix} A & 0 \\ 0 & (A^{-1})^T \end{pmatrix},
\end{equation}
where $A\in\GL(d,\bbR)$ is the matrix transforming $\hat{e}_a\mapsto
\hat{e}_b(A^{-1})^b{}_a$ while $\omega=\frac{1}{2}\omega_{ab}e^a\wedge
e^b$ transforms $B\mapsto B'=B+\omega$, where $\omega$ must be closed for
$B'$ to be a splitting. This defines a parabolic subgroup
$\Ggeom=\GL(d,\bbR)\ltimes\bbR^{d(d-1)/2}\subset O(d,d)\times\bbR^+$
and hence the set of all frames of the form~\eqref{eq:geom-basis}
defines a $\Ggeom$ principal sub-bundle of $\tilde{F}$, that is a
$\Ggeom$-structure. This reflects the fact that the patching elements
in the definition of $\tilde{E}$ lie only in this subgroup of
$O(d,d)\times\bbR^+$.

In what follows it will be useful to also define a class of
\emph{conformal split frames} given by the set of split bases 
conformally rescaled by a function $\phi$ so that 
\begin{equation}
\label{eq:Csplit}
   \hat{E}_A = \begin{cases} 
           \hat{E}_a = \ee^{-2\phi}(\det e) \left(
              \hat{e}_a + i_{\hat{e}_a} B \right) 
               & \text{for $A=a$} \\
           E^a = \ee^{-2\phi}(\det e) e^a &  \text{for $A=a+d$}
         \end{cases} . 
\end{equation}
thus defining a $\Ggeom\times\bbR^+$ sub-bundle of $\tilde{F}$. In
complete analogy with the split case, the components of
$V\in\Gs{\tilde{E}}$ in the conformally split frame are related to
those in the coordinate basis by 
\begin{equation}
   V^{(B,\phi)} 
     = \ee^{2\phi}\big(v_{(i)} + \lambda_{(i)} - i_{v_{(i)}} B_{(i)}\big).
\end{equation}

We can similarly write the components of generalised spinors in
different frames. The relation between the coordinate and split frames
implies that if $\Psi^{(B)}_{a_1\dots a_n}$ are the polyform
components of $\Psi\in\Gs{S^\pm_{(p)}}$ in the split frame then 
\begin{equation}
\label{eq:PsiB}
   \Psi^{(B)} 
     = \sum_n \tfrac{1}{n!}\Psi^{(B)}_{a_1\dots a_n}
         e^{a_1}\wedge\dots\wedge e^{a_n}
     = \ee^{B_{(i)}} \wedge \Psi_{(i)} , 
\end{equation}
demonstrating the isomorphism
$S^\pm_{(p)}\simeq(\det{T^*M})^{p-1/2}\otimes\Leo T^*M$, since
$\ee^{B_{(i)}} \wedge \Psi_{(i)} = \ee^{B_{(j)}} \wedge
\Psi_{(j)}$. In the conformal split frame one similarly has 
\begin{equation}
   \Psi^{(B, \phi)} = \ee^{p\phi} \ee^{B_{(i)}} \wedge \Psi_{(i)} . 
\end{equation}
%


\subsection{The Dorfman derivative, Courant bracket and exterior derivative}
\label{sec:gen-lie}

An important property of the generalised tangent space is that it
admits a generalisation of the Lie derivative which encodes the
bosonic symmetries of the NSNS sector of type II supergravity. Given
$V=v+\lambda\in\Gs{E}$, one can define an operator $\Lgen_V$ acting on
any generalised tensor, which combines the action of an infinitesimal
diffeomorphisms generated by $v$ and a $B$-field gauge transformations
generated by $\lambda$. 

Acting on $W=w+\zeta\in E_{(p)}$, we define the \emph{Dorfman
derivative}\footnote{If $p=0$ then $\Lgen_VW$ is none other than the
  Dorfman bracket~\cite{dorfman}. Since it extends to a derivation on
  the tensor algebra of generalised tensors, it is natural in our
  context to call it the ``Dorfman derivative''.} or ``generalised Lie
derivative'' as~\cite{GMPW} 
\begin{equation}
\label{eq:Lgen}
   \Lgen_V W = \mathcal{L}_v w + \mathcal{L}_v \zeta - i_w \dd\lambda ,
\end{equation}
where, since $w$ and $\zeta$ are weighted tensors, the action of the Lie
derivative is
\begin{equation}
\begin{aligned}
   \mathcal{L}_v w^\mu 
      &= v^\nu \der_\nu w^\mu - w^\nu\der_\nu v^\mu + p(\der_\nu v^\nu) w^\mu , \\
   \mathcal{L}_v \zeta_\mu 
      &= v^\nu\der_\nu \zeta_\mu + (\der_\mu v^\nu)\zeta_\nu + p(\der_\nu v^\nu) \zeta^\mu . 
\end{aligned}
\end{equation}
Defining the action on a function $f$ as simply $\Lgen_V
f=\mathcal{L}_vf$, one can then extend the notion of Dorfman
derivative to  any $O(d,d)\times\bbR^+$ tensor using the Leibniz
property. 

To see this explicitly it is useful to note that we can
rewrite~\eqref{eq:Lgen} in a more $O(d,d)\times\bbR^+$ covariant way,
in analogy with~\eqref{eq:LDgen}. First note that one can embed the
action of the partial derivative operator into generalised
geometry using the map $T^*M\to E$. In coordinate indices, as viewed
as mapping to a section of $E^*$, one defines   
\begin{equation}
\label{eq:d-def}
   \der_M 
      = \begin{cases} \der_\mu  & \text{for $M=\mu$} \\
         0 & \text{for $M=\mu+d$}
         \end{cases} . 
\end{equation}
One can then rewrite~\eqref{eq:Lgen} in terms of generalised
objects (as in~\cite{Siegel:1993xq,Siegel:1993th,Hohm:2010pp})
\begin{equation}
\label{eq:Lgen-cov}
   \Lgen_V W^M
      = V^N \der_N W^M + \left(\der^M V^N - \der^N V^M\right) W_N
         + p \left(\der_N V^N\right) W^M , 
\end{equation}
where indices are contracted using the $O(d,d)$
metric~\eqref{eq:Odd-metric}, which, by definition, is constant with
respect to $\der$. Note that this form is exactly analogous to the
conventional Lie derivative~\eqref{eq:LDgen}, though now with the
adjoint action in $\mathfrak{o}(d,d)\oplus \bbR$ rather than 
$\mathfrak{gl}(d)$. Specifically the second and third terms are (minus) the
action of an $\mathfrak{o}(d,d)\oplus \bbR$ element $m$, given by 
\begin{equation}
   m \cdot W
     = \begin{pmatrix} a & 0 \\ -\omega & -a^T \end{pmatrix}
       \begin{pmatrix} w \\ \zeta \end{pmatrix}
       - p \tr a \begin{pmatrix} w \\ \zeta \end{pmatrix} , 
\end{equation}
where $a^\mu{}_\nu=\der_\nu v^\mu$ and
$\omega_{\mu \nu}=\der_\mu\lambda_\nu-\der_\nu \lambda_\mu$. Comparing
with~\eqref{eq:Ggeom}, we see that $m$ in fact acts in the Lie algebra of the
$\Ggeom$ subgroup of $O(d,d)\times\bbR^+$. 

This form can then be naturally extended to an arbitrary
$O(d,d)\times\bbR^+$ tensor $\alpha\in\Gs{E^{\otimes n}_{(p)}}$ as 
\begin{equation}
\label{eq:Lgen-gen}
\begin{aligned}
   \Lgen_V \alpha^{M_1\dots M_n}
      &= V^N \der_N \alpha^{M_1\dots M_n} 
         + \left(\der^{M_1} V^N - \der^N V^{M_1}\right) 
             \alpha_N{}^{M_2\dots M_n} 
         \\ & \qquad
         + \dots 
         + \left(\der^{M_n} V^N - \der^N V^{M_n}\right) 
             \alpha^{M_1\dots M_{n-1}}{}_N
         + p \left(\der_N V^N\right) W^M , 
\end{aligned}
\end{equation}
again in analogy with~\eqref{eq:LDgen}. It similarly extends to generalised
spinors $\Psi\in \Gs{S^\pm_{(p)}}$ as (see also~\cite{Hohm:2011zr})
\begin{equation}
   \Lgen_V \Psi
      = V^N \der_N \Psi + \tfrac{1}{4}\left(
            \der_M V_N - \der_N V_M\right)\Gamma^{MN}\Psi
            + p(\der_M V^M) \Psi , 
\end{equation}
where
$\Gamma_{MN}=\frac{1}{2}\left(\Gamma_M\Gamma_N-\Gamma_N\Gamma_M\right)$. 

Note that when $W\in \Gs{E}$ one can also define the antisymmetrisation of
the Dorfman derivative  
\begin{equation}
\label{eq:Courant}
\begin{aligned}
   \Bgen{V}{W} &= \tfrac{1}{2}\left( \Lgen_V W - \Lgen_W V \right) \\
      &= \BLie{v}{w} + \mathcal{L}_v \zeta - \mathcal{L}_w \lambda
         - \tfrac{1}{2}\dd\left(i_v\zeta - i_w\lambda\right) , 
\end{aligned}
\end{equation}
which is known as the Courant bracket~\cite{courant}. It can be
rewritten in an $O(d,d)$ covariant form as 
\begin{equation}
\label{eq:Courant-cov}
   \Bgen{U}{V}^M 
      = U^N \der_N V^M - V^N \der_N U^M 
         - \tfrac{1}{2}\left(U_N\der^M V^N - V_N\der^M U^N\right) . 
\end{equation}
which follows directly from~\eqref{eq:Lgen-cov}. 

Finally note that since $S^\pm_{(1/2)}\simeq\Leo T^*M$ the Clifford
action of $\der_M$ on $\Psi\in\Gs{S^\pm_{(1/2)}}$ defines a natural
action of the exterior derivative. On $U_i$ one defines
$d:\Gs{S^\pm_{(1/2)}}\to\Gs{S^\mp_{(1/2)}}$ by 
\begin{equation}
\label{eq:gen-dd}
   \left( \dd \Psi\right)_{(i)} 
      = \tfrac{1}{2}\Gamma^M\der_M \Psi_{(i)}
      = \dd \Psi_{(i)} ,
\end{equation}
that is, it is simply the exterior derivative of the component
$p$-forms. The Dorfman derivative and Courant bracket can then be
regarded as derived brackets for this exterior derivative~\cite{K-S}.


\subsection{Generalised $O(d,d)\times\bbR^+$ connections and torsion} 
\label{sec:conns}

We now turn to the definitions of generalised connections, torsion and
the possibility of defining a generalised curvature. The notion of
connection on a Courant algebroid was first introduced by Alekseev and
Xu~\cite{AX,CSX} and Gualtieri~\cite{gualtieri-conn} (see also
Ellwood~\cite{ellwood}). At least locally, it is also essentially
equivalent to the connection defined by
Siegel~\cite{Siegel:1993xq,Siegel:1993th} and discussed in double
field theory~\cite{Hohm:2010xe}. It is also very closely related to
the differential operator introduced in the ``stringy
differential geometry'' of~\cite{JLP}. 

Our definitions will follow closely those in~\cite{AX,gualtieri-conn}
though, in connecting to supergravity, it is important to extend the
definitions to include the $\bbR^+$ factor in the generalised
structure bundle. 


\subsubsection{Generalised connections}
\label{sec:conn-def}

Here we will specifically be interested in those generalised
connections that are compatible with the $O(d,d)\times\bbR^+$
structure. Following~\cite{AX,gualtieri-conn} we can define a
first-order linear differential operator $\Dgen$, such that, given 
$W\in\Gs{\tilde{E}}$, in frame indices, 
\begin{equation}
   \Dgen_M W^A = \der_M W^A + \tilde{\Omega}_M{}^A{}_B W^B . 
\end{equation}
Compatibility with the $O(d,d)\times\bbR^+$ structure implies
\begin{equation}
   \tilde{\Omega}_M{}^A{}_B 
      = \Omega_M{}^A{}_B - \Lambda_M \delta^A{}_B , 
\end{equation}
where $\Lambda$ is the $\bbR^+$ part of the connection and $\Omega$
the $O(d,d)$ part, so that we have 
\begin{equation}
   \Omega_M{}^{AB} = - \Omega_M{}^{BA} . 
\end{equation}
The action of $\Dgen$ then extends naturally to any generalised
tensor. In particular, if $\alpha\in\Gs{E^{\otimes n}_{(p)}}$ we have  
\begin{equation}
 \begin{aligned}
   \Dgen_M \alpha^{A_1\dots A_n}
      &= \der_M \alpha^{A_1\dots A_n} 
          + \Omega_M{}^{A_1}{}_B \alpha^{B A_2\dots A_n} 
          \\ & \qquad \qquad 
          + \dots + \Omega_M{}^{A_n}{}_B \alpha^{A_1\dots A_{n-1} B} 
          - p \Lambda_M \alpha^{A_1\dots A_n} . 
\end{aligned}
\end{equation}
Similarly, if $\Psi\in\Gs{S^\pm_{(p)}}$ then 
\begin{equation}
   \Dgen_M \Psi = \left(\der_M 
      + \tfrac{1}{4}\Omega_M{}^{AB}\Gamma_{AB}
      - p \Lambda_M \right) \Psi .  
\end{equation}

Given a conventional connection $\nabla$ and a conformal split frame
of the form~\eqref{eq:Csplit}, one can construct the corresponding
generalised connection as follows. Writing a generalised vector
$W\in\Gs{\tilde{E}}$ as 
\begin{equation}
   W = W^A\hat{E}_A = w^a \hat{E}_a + \zeta_a E^a ,
\end{equation}
by construction 
$w=w^a (\det{e})\hat{e}_a\in\Gs{(\det{T^*M})\otimes TM}$ and 
$\zeta=\zeta_a (\det{e})e^a \in\Gs{(\det{T^*M})\otimes T^*M}$ and
so we can define $\nabla_\mu w^a$ and $\nabla_\mu\zeta_a$. The generalised
connection defined by $\nabla$ lifted to an action on $\tilde{E}$ by
the conformal split frame is then simply 
\begin{equation}
\label{eq:Dgen-embed}
   (\Dgen^\nabla_M W^A) \hat{E}_A= \begin{cases}
      (\nabla_\mu w^a) \hat{E}_a + (\nabla_\mu \zeta_a) E^a  
      & \text{for $M=\mu$} \\
      0  & \text{for $M=\mu+d$} 
      \end{cases} . 
\end{equation}
%


\subsubsection{Generalised torsion}
\label{sec:torsion}

We define the \emph{generalised torsion} $T$ of a generalised connection
$\Dgen$ in direct analogy to the conventional
definition~\eqref{eq:Tdef2}. Let $\alpha$ be any generalised tensor
and $\Lgen^\Dgen_V\alpha$ be the Dorfman
derivative~\eqref{eq:Lgen-gen} with $\der$ replaced by $\Dgen$. The
generalised torsion is a linear map $T:\Gs{E}\to\Gs{\adj(\tilde{F})}$
where $\adj(\tilde{F})\simeq\Lambda^2E\oplus\bbR$ is the $\mathfrak{o}(d,d)
\oplus \bbR$ adjoint representation bundle associated to
$\tilde{F}$. It is defined by 
\begin{equation}
\label{eq:Tgen-def}
   T(V)\cdot \alpha 
       = \Lgen^\Dgen_V \alpha - \Lgen_V \alpha , 
\end{equation}
for any $V\in\Gs{E}$ and where $T(V)$ acts via the adjoint
representation on $\alpha$. This definition is close to that
of~\cite{gualtieri-conn}, except for the additional $\bbR^+$
action in the definition of $\Lgen$. 

Viewed as a tensor $T\in\Gs{E\otimes\adj{\tilde{F}}}$, with indices
such that $T(V)^M{}_N=V^P T^M{}_{PN}$, we can derive an explicit
expression for $T$. Let $\{\hat{E}_A\}$ be a general conformal basis
with $\GM{\hat{E}_A}{\hat{E}_B}=\Phi^2\eta_{AB}$. Then
$\{\Phi^{-1}\hat{E}_A\}$ is an orthonormal basis for $E$. Given 
the connection $\Dgen_M W^A=\der_M W^A+\tilde{\Omega}_M{}^A{}_B W^B$,
we have  
\begin{equation}
\label{eq:Tgen-basis}
   T_{ABC} = -3\tilde{\Omega}_{[ABC]} 
         + \tilde{\Omega}_D{}^D{}_B \eta_{AC} 
         - \Phi^{-2}\GM{\hat{E}_A}{\Lgen_{\Phi^{-1}\hat{E}_B}\hat{E}_C} , 
\end{equation}
where indices are lowered with $\eta_{AB}$. 

Naively one might expect that $T\in\Gs{(E\otimes\Lambda^2E)\oplus
  E}$. However the form of the Dorfman derivative means that fewer
components of $\tilde{\Omega}$ actually enter the torsion and 
\begin{equation}
   T \in \Gs{\Lambda^3 E \oplus E} . 
\end{equation}
This can be seen most easily in the coordinate basis where the two
components are 
\begin{equation}
\label{eq:TAV}
   T^{M}{}_{PN} = (\TA)^M{}_{PN} - (\TV){}_P \,\delta^M{}_N , 
\end{equation}
with
\begin{equation}
\label{eq:Tgen-comp}
\begin{aligned}
   (\TA){}_{MNP} &= -3\tilde{\Omega}_{[MNP]} = -3\Omega_{[MNP]} , \\
   (\TV){}_M &= -\tilde{\Omega}_Q{}^Q{}_M
         =  \Lambda_M - \Omega_Q{}^Q{}_M .
\end{aligned}
\end{equation}

An immediate consequence of this definition is that for $\Psi \in
\Gs{S^\pm_{(1/2)}}$ the Dirac operator $\Gamma^M \Dgen_M \Psi$ is
determined by the torsion of the connection~\cite{AX}
\begin{equation}
\begin{split}
   \Gamma^M \Dgen_M \Psi &= \Gamma^M (\der_M \Psi 
         + \tfrac14 \Omega_{MNP} \Gamma^{NP} \Psi
         - \tfrac12 \Lambda_M \Psi ) \\
      &= \Gamma^M \der_M \Psi + \tfrac14 \Omega_{[MNP]} \Gamma^{MNP} \Psi
         - \tfrac12 (\Lambda_M - \Omega_N{}^N{}_M ) \Gamma^M \Psi \\
      &= 2\dd\Psi - \tfrac{1}{12} (\TA)_{[MNP]} \Gamma^{MNP} \Psi
         - \tfrac12 (\TV)_M \Gamma^M \Psi .
\end{split}
\end{equation}
This equation could equally well be used as a definition of the
torsion of a generalised connection. Note in particular that if the
connection is torsion-free we see that the Dirac operator becomes
equal to the exterior derivative 
\begin{equation}
\label{eq:gen-dd-torsion}
   \Gamma^M \Dgen_M \Psi = 2 \dd \Psi . 
\end{equation}

As an example, we can calculate the torsion for the generalised connection
$\Dgen^\nabla$ defined in~\eqref{eq:Dgen-embed}. In general we have 
\begin{equation}
\label{eq:B-embed1}
   \Lgen_{\Phi^{-1}\hat{E}_A}\hat{E}_B
      = \left(\Lgen_{\Phi^{-1}\hat{E}_A}\Phi\right)\Phi^{-1}\hat{E}_B
          + \Phi\big(\Lgen_{\Phi^{-1}\hat{E}_A}
             (\Phi^{-1}\hat{E}_B) \big) , 
\end{equation}
where here
\begin{equation}
\label{eq:B-embed1}
   \Lgen_{\Phi^{-1}\hat{E}_A}\Phi
      = \begin{cases} 
            -\ee^{-2\phi}(\det e) \left( 
               i_{\hat{e}_a}i_{\hat{e}_b}\dd e^b 
               + 2 i_{\hat{e}_a}\dd\phi \right) 
            & \text{for $A=a$} \\
            0 & \text{for $A=a+d$}
        \end{cases} ,
\end{equation}
and
\begin{equation}
\label{eq:B-embed2}
   \Lgen_{\Phi^{-1}\hat{E}_A}\Phi^{-1}\hat{E}_B
      = \begin{pmatrix}
            \BLie{\hat{e}_a}{\hat{e}_b} 
               + i_{\BLie{\hat{e}_a}{\hat{e}_b}}B 
               - i_{\hat{e}_a}i_{\hat{e}_b}H \quad  & 
            \mathcal{L}_{\hat{e}_a}e^b \\ 
            -\mathcal{L}_{\hat{e}_b}e^a &  
            0
        \end{pmatrix}_{AB} ,
\end{equation}
where $H=\dd B$. If the conventional connection $\nabla$ is torsion-free, the
corresponding generalised torsion is given by
\begin{equation}
\label{eq:Hphi-tor}
   \TA = - 4 H , \qquad
   \TV = - 4 \, \dd \phi , 
\end{equation}
where we are using the embedding\footnote{Note that with our definitions we have $ (\der^A \phi) \Phi^{-1} \hat{E}_A = 2 \dd \phi$ due to the factor $\tfrac12$ in $\eta_{AB}$}
 $T^*M\to E$ (and the corresponding
$T^*M\to\Lambda^3E$) to write the expressions in terms of forms. This
result is most easily seen by taking $\hat{e}_a$ to be the coordinate
frame, so that all but the $H$ and $\dd\phi$ terms
in~\eqref{eq:B-embed1} and~\eqref{eq:B-embed2} vanish.  


\subsubsection{The absence of generalised curvature}
\label{sec:gen-curv}

Having defined torsion it is natural to ask if one can also introduce
a notion of generalised curvature in analogy to the usual definition~\eqref{eq:Rdef},
as the commutator of two generalised connections but now using the
Courant bracket~\eqref{eq:Courant} rather than the Lie
bracket 
\begin{equation}
\label{eq:Dgen-comm}
\GenR \left(U,V,W\right) = \BLie{\Dgen_U}{\Dgen_V}W - \Dgen_{\Bgen{U}{V}} W .
\end{equation}
However, this object is non-tensorial~\cite{gualtieri-conn}. We can
check for linearity in the arguments explicitly. Taking $U \rightarrow
fU$, $V \rightarrow gV$ and  $W \rightarrow hW$ for some scalar
functions $f,g,h$, we obtain 
\begin{equation}
\label{eq:Dgen-comm-linear}
\begin{aligned}
   &\BLie{\Dgen_{fU}}{\Dgen_{gV}}hW - \Dgen_{\Bgen{fU}{gV}} hW 
      \\ & \qquad 
      = fgh\left( \BLie{\Dgen_U}{\Dgen_V}W - \Dgen_{\Bgen{U}{V}} W \right)
                - \tfrac{1}{2} h \GM{U}{V} \Dgen_{\left(f\dd g - g \dd f\right)} W , 
\end{aligned}
\end{equation}
and so the curvature is not linear in $U$ and $V$. 

Nonetheless, if there is additional structure, as will be relevant for
supergravity, we are able to define other tensorial objects that are
measures of generalised curvature. In particular, let $C_1\subset E$
and $C_2\subset E$ be subspaces such that $\GM{U}{V}=0$ for all
$U\in\Gs{C_1}$ and $V\in\Gs{C_2}$. For such a $U$ and $V$ the final
term in~\eqref{eq:Dgen-comm-linear} vanishes, and so $\GenR \in
\Gs{\left( C_1 \otimes C_2 \right) \otimes \mathfrak{o}(d,d)}$ is a tensor.  
A special example of this is when $C_1=C_2$ is a null subspace
of $E$. 


\section{$O(p,q)\times O(q,p)$ structures and torsion-free connections}
\label{sec:OdOd-sec}

We now turn to constructing the generalised analogue of the
Levi--Civita connection. The latter is the unique torsion-free
connection that preserves the $O(d)\subset\GL(d,\bbR)$ structure
defined by a metric $g$. Here we will be interested in generalised
connections that preserve an $O(p,q)\times O(q,p)\subset
O(d,d)\times\bbR^+$ structure on $\tilde{F}$, where $p+q=d$. We will
find that, in analogy to the Levi--Civita connection, it is always
possible to construct torsion-free connections of this type but
there is no unique choice. Locally this is same construction that
appears in Siegel~\cite{Siegel:1993xq,Siegel:1993th} and closely
related to that of~\cite{JLP}.


\subsection{$O(p,q)\times O(q,p)$ structures and the generalised metric}
\label{sec:OdOd}

Following closely the standard definition of the generalised
metric~\cite{Gualtieri}, consider an $O(p,q)\times O(q,p)$ principal
sub-bundle $P$ of the generalised structure bundle $\tilde{F}$. As
discussed below, this is equivalent to specifying a conventional
metric $g$ of signature $(p,q)$, a $B$-field patched as
in~\eqref{eq:Bpatch} and a dilaton $\phi$. As such it clearly gives
the appropriate generalised structure to capture the NSNS supergravity
fields. 

Geometrically, an $O(p,q)\times O(q,p)$ structure does two
things. First it fixes a nowhere vanishing section
$\Phi\in\Gs{\det T^*M}$, giving an isomorphism between
weighted and unweighted generalised tangent space $\tilde{E}$ and
$E$. Second it defines a splitting of $E$ into two $d$-dimensional
sub-bundles  
\begin{equation}
\label{eq:C+C-}
   E = C_+ \oplus C_- \, ,
\end{equation}
such that the $O(d,d)$ metric~\eqref{eq:Odd-metric} restricts to a
separate metric of signature $(p,q)$ on $C_+$ and a metric of
signature $(q,p)$ on $C_-$. (Each sub-bundle is also isomorphic to
$TM$ using the map $E\to TM$.)

In terms of $\tilde{F}$ we can identify a special set of frames
defining a $O(p,q)\times O(p,q)$ sub-bundle. We define a frame
$\{\hat{E}^+_a\}\cup\{\hat{E}^-_{\bar{a}}\}$ such that
$\{\hat{E}^+_a\}$ form an orthonormal basis for $C_+$ and
$\{\hat{E}^-_{\bar{a}}\}$ form an orthonormal basis for 
$C_-$. This means they satisfy  
\begin{equation}
\label{eq:OdOdframe}
\begin{aligned}
   \GM{\hat{E}^+_a}{\hat{E}^+_b} &= \Phi^2 \eta_{ab} , \\
   \GM{\hat{E}^-_{\bar{a}}}{\hat{E}^-_{\bar{b}}}
       &= - \Phi^2 \eta_{\bar{a}\bar{b}} ,\\
   \GM{\hat{E}^+_a}{\hat{E}^-_{\bar{a}}} &= 0 . 
\end{aligned}
\end{equation}
where $\Phi\in\Gs{\det T^*M}$ is now some fixed density (independent
of the particular frame element) and $\eta_{ab}$ and
$\eta_{\bar{a}\bar{b}}$ are flat metrics with signature $(p,q)$. There
is thus a manifest $O(p,q)\times O(q,p)$ symmetry with the first
factor acting on $\hat{E}^+_a$ and the second on
$\hat{E}^-_{\bar{a}}$.

Note that the natural conformal frame 
\begin{equation}
   \hat{E}_A = 
      \begin{cases} 
         \hat{E}^+_a & \text{for $A=a$} \\
         \hat{E}^-_{\bar{a}} &  \text{for $A=\bar{a}+d$}
      \end{cases} , 
\end{equation}
satisfies 
\begin{equation}
   \GM{\hat{E}_A}{\hat{E}_B} = \Phi^2 \eta_{AB} ,
   \quad \text{where} \quad 
   \eta_{AB} = \begin{pmatrix} 
      \eta_{ab} & 0 \\ 0 & -\eta_{\bar{a} \bar{b}} \end{pmatrix} ,
\end{equation}
where the form of $\eta_{AB}$ differs from that used
in~\eqref{eq:conformal}. In this section, we will use this form of the
metric $\eta_{AB}$ throughout. It is also important to note that we
will adopt the convention that we will always raise and lower the
$C_+$ indices $a,b,c,\dots $ with $\eta_{ab}$ and the $C_-$ indices
$\bar{a},\bar{b},\bar{c},\dots$ with $\eta_{\bar{a} \bar{b}}$, while
we continue to raise and lower $2d$ dimensional indices $A,B,C,\dots$
with the $O(d,d)$ metric $\eta_{AB}$. Thus, for example we have
\begin{equation}
   \hat{E}^A = 
      \begin{cases} 
         \hat{E}^{+a} & \text{for $A=a$} \\
         - \hat{E}^{-\bar{a}} &  \text{for $A=\bar{a}+d$}
      \end{cases} , 
\end{equation}
when we raise the $A$ index on the frame. 

One can write a generic $O(p,q)\times O(q,p)$ structure explicitly as 
\begin{equation}
\label{eq:OdOdexplicit}
\begin{aligned}
   \hat{E}^+_a &= \ee^{-2\phi}\sqrt{-g} 
      \left( \hat{e}^+_a + e^+_a + i_{\hat{e}^+_a}B \right) , \\
   \hat{E}^-_{\bar{a}} &= \ee^{-2\phi}\sqrt{-g} 
      \left( \hat{e}^-_{\bar{a}} - e^-_{\bar{a}} + i_{\hat{e}^-_{\bar{a}}}B \right) , 
\end{aligned}
\end{equation}
where the fixed conformal factor in~\eqref{eq:OdOdframe} is given by 
\begin{equation}
\label{eq:gendilaton}
   \Phi = \ee^{-2\phi}\sqrt{-g} ,
\end{equation}
and where $\{\hat{e}^+_a\}$ and $\{\hat{e}^-_{\bar{a}}\}$, and their
duals $\{e^{+a}\}$ and $\{e^{-\bar{a}}\}$, are two independent
orthonormal frames for the metric $g$, so that 
\begin{equation}
\begin{gathered}
   g = \eta_{ab} e^{+a}\otimes e^{+b} 
      = \eta_{\bar{a}\bar{b}}e^{-\bar{a}}\otimes e^{-\bar{b}} , \\
   g(\hat{e}^+_a,\hat{e}^+_b) = \eta_{ab} , \qquad
   g(\hat{e}^-_{\bar{a}},\hat{e}^-_{\bar{b}}) 
      = \eta_{\bar{a}\bar{b}} .
\end{gathered}
\end{equation}
By this explicit construction we see that there is no topological
obstruction to the existence of $O(p,q)\times O(q,p)$ structures. 

In addition to the $O(p,q)\times O(q,p)$ invariant
density~\eqref{eq:gendilaton} one can also construct the invariant 
\emph{generalised metric}  $G$~\cite{Gualtieri}. It has the form
\begin{equation}
   G = \Phi^{-2}\big(\eta^{ab} \hat{E}^+_a\otimes\hat{E}^+_b
        + \eta^{\bar{a}\bar{b}} 
           \hat{E}^-_{\bar{a}}\otimes\hat{E}^-_{\bar{b}} \big) .
\end{equation}
In the coordinate frame we have the familiar expression
\begin{equation}
   G_{MN} = \frac12 \begin{pmatrix}
	  	g - Bg^{-1}B & -Bg^{-1} \\
		g^{-1}B & g^{-1} 
             \end{pmatrix}_{MN} . 
\end{equation}
By construction, the pair $(G,\Phi)$ parametrise the coset
$(O(d,d)\times\bbR^+)/O(p,q)\times O(q,p)$ where $p+q=d$.

Finally the $O(p,q)\times O(q,p)$ structure provides two additional
chirality operators $\Gamma^\pm$ on $\Spin(d,d)\times\mathbb{R}^+$
spinors which one can define
as~\cite{GMPW,gen-hodge,Hohm:2011zr}  
\begin{equation}
\label{eq:OdxOd-chirality}
   \Gamma^{(+)} = \tfrac{1}{d!} 
      \epsilon^{a_1 \dots a_d} \Gamma_{a_1} \dots \Gamma_{a_d} ,\qquad \qquad
   \Gamma^{(-)} = \tfrac{1}{d!} 
      \epsilon^{\bar{a}_1 \dots \bar{a}_d} 
      \Gamma_{\bar{a}_1} \dots \Gamma_{\bar{a}_d} .
\end{equation}
Using that, in the split frame, the Clifford action takes the form
\begin{equation}
   \Gamma_a \cdot \Psi^{(B)} 
      = i_{\hat{e}^+_a} \Psi^{(B)} + e^+_a \wedge \Psi^{(B)}, \qquad
   \Gamma_{\bar{a}} \cdot \Psi^{(B)} 
      = i_{\hat{e}^-_a} \Psi^{(B)} - e^-_a \wedge \Psi^{(B)} ,
\end{equation}
these can be evaluated on the weighted n-form components of $\Psi$ as
\begin{equation}
   \Gamma^{(+)} \Psi^{(B)}_{(n)} 
      = (-)^{[n/2]} * \Psi^{(B)}_{(n)} ,\qquad \qquad
   \Gamma^{(-)} \Psi^{(B)}_{(n)} 
      = (-)^d (-)^{[n+1/2]} * \Psi^{(B)}_{(n)} ,
\end{equation}
and thus we have a generalisation of the Hodge dual on
$\Spin(d,d)\times\mathbb{R}^+$ spinors. 

Since $G^T \eta G = \eta$, the generalised metric $G^A{}_B$ is an
element of $O(d,d)$ and one can easily check that $G^2 =
\id$. Connecting to the discussion of~\cite{Hohm:2011zr},
for even dimensions $d$, one has $G \in SO(d,d)$ and $\Gamma^{(-)} $
is an element of $\Spin(d,d)$ satisfying 
\begin{equation}
	\Gamma^{(-)} \Gamma^A \Gamma^{(-)-1} = G^A {}_B \Gamma^B ,
\end{equation}
so that $\Gamma^{(-)}$ is a preimage of $G$ in the double covering map
$\Spin(d,d) \rightarrow SO(d,d)$. In odd dimensions $d$, $\Gamma^{(+)}$
is an element of $Pin(d,d)$ which maps to $G \in O(d,d)$ under the
double cover $Pin(d,d) \rightarrow O(d,d)$.


\subsection{Torsion-free, compatible connections}
\label{sec:genLC}

A generalised connection $\Dgen$ is compatible with the $O(p,q)\times
O(q,p)$ structure $P\subset\tilde{F}$ if 
\begin{equation}
   \Dgen G = 0 , \qquad \Dgen \Phi = 0 , 
\end{equation}
or equivalently, if the derivative acts only in the $O(p,q)\times O(q,p)$
sub-bundle so that for $W\in\Gs{\tilde{E}}$ given by 
\begin{equation}
\label{eq:W-pm-basis}
   W = w_+^a \hat{E}^+_a + w_-^{\bar{a}} \hat{E}^-_{\bar{a}} , 
\end{equation}
we have
\begin{equation}
	D_M W^A = \begin{cases} 
		\der_M w^a_+ + \Omega_M{}^a{}_b w^b_+ \qquad \text{for } A = a \\
		\der_M w^{\bar{a}}_- + \Omega_M{}^{\bar{a}}{}_{\bar{b}} w^{\bar{b}} _- \qquad \text{for } A = \bar{a}
	\end{cases} ,
\end{equation}
with
\begin{equation}
   \Omega_{Mab} = - \Omega_{Mba} , \qquad
   \Omega_{M\bar{a}\bar{b}} = - \Omega_{M\bar{b}\bar{a}} . 
\end{equation}
In this subsection we will show, in analogy to the construction of the
Levi--Civita connection, that
\begin{quote}
   \textit{Given an $O(p,q)\times O(q,p)$ structure $P\subset\tilde{F}$
     there always exists a torsion-free, compatible generalised
     connection $\Dgen$. However, it is not unique.} 
\end{quote}

We can construct a compatible connection as follows. Let $\nabla$ be
the Levi--Civita connection for the metric $g$. In terms of the two
orthonormal bases we get two gauge equivalent spin-connections, so
that if $v=v^a\hat{e}^+_a=v^{\bar{a}}\hat{e}^-_{\bar{a}}\in\Gs{TM}$ we have   
\begin{equation}
   \nabla_\mu v^\nu 
      = \big( \der_\mu v^a 
         + \omega^+_\mu{}^a{}_b v^b \big) (\hat{e}^+_a){}^\nu
      = \big( \der_\mu v^{\bar{a}} 
         + \omega^-_\mu{}^{\bar{a}}{}_{\bar{b}} v^{\bar{b}} 
         \big) (\hat{e}^-_{\bar{a}}){}^\nu . 
\end{equation}
We can then define, as in~\eqref{eq:Dgen-embed}
\begin{equation}
\label{eq:Dgen-LC}
   \Dgen^\nabla_M W^a 
      = \begin{cases} 
         \LC_\mu w^a_+ & \text{for } M = \mu \\
         0 & \text{for } M = \mu + d \end{cases}, \qquad
	\Dgen^\nabla_M W^{\bar{a}} = \begin{cases} \LC_\mu
           w^{\bar{a}}_- & 
           \text{for } M = \mu \\
           0 & \text{for } M = \mu + d 
        \end{cases}.
\end{equation}
Since $\omega^+_{\mu ab}=-\omega^+_{\mu ba}$ and
$\omega^-_{\mu\bar{a}\bar{b}}=-\omega^-_{\mu\bar{b}\bar{a}}$, by
construction, this generalised connection is compatible with the
$O(p,q)\times O(q,p)$ structure. 

However $\Dgen^\nabla$ is not torsion-free. To see this we note that,
comparing with~\eqref{eq:Csplit}, when we choose the two orthonormal
frames to be aligned so $e_a^+=e_a^-=e_a$ we have   
\begin{equation}
   W = w_+^a \hat{E}^+_a + w_-^{\bar{a}} \hat{E}^-_{\bar{a}} 
     = \left(w_+^a+w_-^a\right) \hat{E}_a 
         + \left(w_{+a}-w_{-a}\right) E^a , 
\end{equation}
and the two definitions of $\Dgen^\nabla$ in~\eqref{eq:Dgen-embed}
and~\eqref{eq:Dgen-LC} agree. Hence from~\eqref{eq:Hphi-tor} we have
the non-zero torsion components 
\begin{equation}
\label{eq:LC-tor}
   \TA = - 4 H , \qquad
   \TV = -4 \dd \phi . 
\end{equation}

To construct a torsion-free compatible connection we simply modify
$\Dgen^\nabla$. A generic generalised connection $\Dgen$ can be always 
be written as 
\begin{equation}
   \Dgen_M W^ A = \Dgen^\nabla_M W^A + \Sigma_M{}^A{}_B W^B . 
\end{equation}
If $\Dgen$ is compatible with the $O(p,q)\times O(q,p)$ structure then
we have $\Sigma_M{}^a{}_{\bar{b}}=\Sigma_M{}^{\bar{a}}{}_b=0$ and 
\begin{equation}
   \Sigma_{Mab} = - \Sigma_{Mba} , \qquad
   \Sigma_{M\bar{a}\bar{b}} = - \Sigma_{M\bar{b}\bar{a}} . 
\end{equation}
By definition, the generalised torsion components of $\Dgen$ are then
given by
\begin{equation}
\label{eq:OdxOd-torsion}
   (\TA)_{ABC} = - 4 H_{ABC} -3 \Sigma_{[ABC]} , \qquad
   (\TV)_A = - 4 \dd \phi_A - \Sigma_C{}^C{}_A . 
\end{equation}
The components $H^{ABC}$ and $\dd\phi^A$ are the components in frame
indices of the corresponding forms under the embeddings
$T^*M\hookrightarrow E$ and
$\Lambda^3T^*M\hookrightarrow\Lambda^3E$. Given  
\begin{equation}
   \dd x^\mu = \tfrac{1}{2}\Phi^{-1}\left( 
      \hat{e}^+_a{}^\mu \hat{E}^{+a} 
         - \hat{e}^-_{\bar{a}}{}^\mu \hat{E}^{-\bar{a}} \right) , 
\end{equation}
we have, for instance, 
\begin{equation}
   \dd\phi = \tfrac{1}{2}\der_a\phi\,\big(\Phi^{-1}\hat{E}^{+a}\big)
       - \tfrac{1}{2}\der_{\bar{a}}\phi\, 
          \big(\Phi^{-1}\hat{E}^{-\bar{a}}\big) . 
\end{equation}
where there is a similar decomposition of $H$ under 
\begin{equation}
   \Lambda^3 T^*M \hookrightarrow \Lambda^3 E 
        \simeq \Lambda^3 C_+ \oplus (\Lambda^2 C_+ \otimes C_-)
        \oplus (C_+\otimes\Lambda^2 C_-) \oplus \Lambda^3 C_- , 
\end{equation} 
Note also that the middle index on $\Sigma_{[ABC]}$ in
equation~\eqref{eq:OdxOd-torsion} has also been lowered with this
$\eta_{AB}$ which introduces some signs. The result is that the
components are 
\begin{equation}
   \dd \phi_A 
      = \begin{cases} 
         \tfrac12 \der_a \phi & A = a \\  
         \tfrac12 \der_{\bar{a}} \phi & A = \bar{a} + d
      \end{cases},
      \qquad 
   H_{ABC} 
      = \begin{cases}
         \tfrac18 H_{abc} & (A,B,C) = (a,b,c) \\
         \tfrac18 H_{ab\bar{c}} & (A,B,C) = (a,b,\bar{c}+d) \\
         \tfrac18 H_{a \bar{b} \bar{c}} & (A,B,C) = (a,\bar{b}+d,\bar{c}+d) \\
         \tfrac18 H_{\bar{a}\bar{b}\bar{c}} & 
            (A,B,C) = (\bar{a}+d,\bar{b}+d,\bar{c}+d)
      \end{cases},
\end{equation}
and that setting the torsion of $\Dgen$ to zero is equivalent to 
\begin{equation}
\label{eq:OddLC}
\begin{aligned}
   \Sigma_{[abc]} &= -\tfrac{1}{6}H_{abc} , & 
   \Sigma_{\bar{a}bc} &= -\tfrac{1}{2}H_{\bar{a}bc} , &
   \Sigma_a{}^a{}_b &= - 2 \der_b \phi , \\
   \Sigma_{[\bar{a}\bar{b}\bar{c}]} 
      &= + \tfrac{1}{6}H_{\bar{a}\bar{b}\bar{c}} , & 
   \Sigma_{a\bar{b}\bar{c}} &= + \tfrac{1}{2}H_{a\bar{b}\bar{c}} , &
   \Sigma_{\bar{a}}{}^{\bar{a}}{}_{\bar{b}} &= - 2\der_{\bar{b}} \phi .
\end{aligned}
\end{equation}
Thus we can always find a torsion-free compatible connection but
clearly these conditions do not determine $\Dgen$
uniquely. Specifically, one finds  
\begin{equation}
\label{eq:Dgen-sol}
\begin{aligned} 
   \Dgen_a w_+^b 
       &= \nabla_a w_+^b - \tfrac{1}{6}H_a{}^b{}_cw_+^c
           - \tfrac{2}{9}\big( 
              \delta_a{}^b \der_c\phi-\eta_{ac}\der^b\phi \big)w_+^c 
           + A^+_a{}^b{}_c w_+^c , \\
   \Dgen_{\bar{a}} w_+^b 
       &= \nabla_{\bar{a}} w_+^b - \tfrac{1}{2}H_{\bar{a}}{}^b{}_cw_+^c , \\ 
   \Dgen_a w_-^{\bar{b}} 
       &= \nabla_a w_-^{\bar{b}} 
           + \tfrac{1}{2}H_a{}^{\bar{b}}{}_{\bar{c}}w_-^{\bar{c}} , \\
   \Dgen_{\bar{a}} w_-^{\bar{b}} 
       &= \nabla_{\bar{a}} w_-^{\bar{b}} 
           + \tfrac{1}{6}H_{\bar{a}}{}^{\bar{b}}{}_{\bar{c}}w_-^{\bar{c}}
           - \tfrac{2}{9}\big( 
              \delta_{\bar{a}}{}^{\bar{b}} \der_{\bar{c}}\phi
              - \eta_{\bar{a}\bar{c}}\der^{\bar{b}}\phi \big)w_-^{\bar{c}} 
           + A^-_{\bar{a}}{}^{\bar{b}}{}_{\bar{c}} w_-^{\bar{c}} , 
\end{aligned}
\end{equation}
where the undetermined tensors $A^\pm$ satisfy 
\begin{equation}
\label{eq:Adef}
\begin{aligned} 
   A^+_{abc}  &= -A^+_{acb} , & 
   A^+_{[abc]} &= 0 , &
   A^+_a{}^a{}_b &= 0 , \\
   A^-_{\bar{a}\bar{b}\bar{c}}  &= -A^-_{\bar{a}\bar{c}\bar{b}}, & 
   A^-_{[\bar{a}\bar{b}\bar{c}]} &= 0 , &
   A^-_{\bar{a}}{}^{\bar{a}}{}_{\bar{b}} &= 0 ,  
\end{aligned}
\end{equation}
and hence do not contribute to the torsion.


\subsection{Unique operators and generalised $O(p,q)\times
  O(q,p)$ curvatures}
\label{sec:curv}

The fact that the $O(p,q)\times O(q,p)$ structure and torsion
conditions are not sufficient to specify a unique generalised
connection might raise ambiguities which could pose a problem for the
applications to supergravity we are ultimately interested in. However,
we will now show that it is still possible to find differential
expressions that are independent of the chosen $\Dgen$, by forming
$O(p,q)\times O(q,p)$ covariant operators which do not depend on the
undetermined components $A^\pm$. For example, by
examining~\eqref{eq:Dgen-sol} we already see that  
\begin{equation}
\label{eq:Dgen-uniq}
\begin{aligned} 
   \Dgen_{\bar{a}} w_+^b 
       &= \nabla_{\bar{a}} w_+^b - \tfrac{1}{2}H_{\bar{a}}{}^b{}_cw_+^c , \\ 
   \Dgen_a w_-^{\bar{b}} 
       &= \nabla_a w_-^{\bar{b}} 
           + \tfrac{1}{2}H_a{}^{\bar{b}}{}_{\bar{c}}w_-^{\bar{c}} , \\
\end{aligned}
\end{equation}
have no dependence on $A^\pm$ and so are unique. We find that this is
also true for 
\begin{equation}
\label{eq:Dgen-v-uniq}
\begin{aligned} 
   \Dgen_a w_+^a &= \nabla_a w_+^a - 2(\der_a\phi) w_+^a, \\
   \Dgen_{\bar{a}} w_-^{\bar{a}} 
      &= \nabla_{\bar{a}} w_-^{\bar{a}} 
         - 2 (\der_{\bar{a}}\phi) w_-^{\bar{a}}.  \\
\end{aligned}
\end{equation}

Anticipating our application to supergravity, we will be especially
interested in writing formulae for $\Spin(p,q)$ spinors, so let us now
assume that we have a $\Spin(p,q)\times Spin(q,p)$ structure. If
$S(C_\pm)$ are then the spinor bundles associated to the sub-bundles
$C_\pm$, $\gamma^a$ and $\gamma^{\bar{a}}$ the corresponding gamma
matrices and $\epsilon^\pm \in \Gs{S(C_\pm)}$, we have that by
definition a generalised connection acts as    
\begin{equation}
\label{eq:spinorDgen0}
\begin{aligned}
   \Dgen_M \epsilon^+ = \der_M \epsilon^+ +
         \tfrac{1}{4}\Omega_M{}^{ab}\gamma_{ab}\epsilon^+ , \\
   \Dgen_M \epsilon^- = \der_M \epsilon^- +
         \tfrac{1}{4}\Omega_M{}^{\bar{a}\bar{b}}
            \gamma_{\bar{a}\bar{b}}\epsilon^- .
\end{aligned} 
\end{equation}
There are four operators which can be built out of these derivatives
that are uniquely determined
\begin{equation}
\label{eq:Dgen-spin-uniq}
\begin{aligned}
   \Dgen_{\bar{a}}\epsilon^+ 
      &= \left( \nabla_{\bar{a}} 
           - \tfrac{1}{8}H_{\bar{a}bc}\gamma^{bc} \right) \epsilon^+ ,
           \\
   \Dgen_a\epsilon^- 
   	&= \left( \nabla_a + \tfrac{1}{8}H_{a\bar{b}\bar{c}}
           \gamma^{\bar{b}\bar{c}} \right) \epsilon^- ,
           \\
   \gamma^a\Dgen_a\epsilon^+ 
      &= \left( \gamma^a\nabla_a - \tfrac{1}{24}H_{abc}\gamma^{abc}
           - \gamma^a\der_a\phi \right) \epsilon^+ , \\
   \gamma^{\bar{a}}\Dgen_{\bar{a}}\epsilon^- 
      &= \left( \gamma^{\bar{a}}\nabla_{\bar{a}} 
           + \tfrac{1}{24}H_{\bar{a}\bar{b}\bar{c}}
               \gamma^{\bar{a}\bar{b}\bar{c}}
           -\gamma^{\bar{a}}\der_{\bar{a}}\phi 
               \right) \epsilon^- . 
\end{aligned}
\end{equation}
The first two expressions follow directly
from~\eqref{eq:Dgen-uniq}. In the final two expressions, there is an
elegant cancellation from
$\gamma^a\gamma^{bc}=\gamma^{abc}+\eta^{ab}\gamma^c-\eta^{ac}\gamma^b$ 
which removes the terms involving $A^\pm$.

The restriction that expressions involving generalised connections be
determined unambiguously, irrespective of the particular $\Dgen$, now
serves as a selection criteria for constructing new generalised
objects. In particular, when defining a generalised notion of
curvature, we find that even though we can actually build a tensorial
$O(p,q)\times O(q,p)$ generalised Riemann curvature -- by
following the example in section~\ref{sec:gen-curv} and taking $C_1 =
C_\pm$ and $C_2 = C_\mp$ so that the index structure would be
$\left(\GenR_{a\bar{b}\phantom{c}d}^{\phantom{ab}c},
\GenR_{a\bar{b}\phantom{\bar{c}}\bar{d}}^{\phantom{ab}\bar{c}}\right)$
and
$\left(\GenR_{\bar{a}b\phantom{c}d}^{\phantom{ab}c},
\GenR_{\bar{a}b\phantom{\bar{c}}\bar{d}}^{\phantom{ab}\bar{c}}\right)$
-- it would not result in a uniquely determined object. However, we
can use combinations of~\eqref{eq:Dgen-uniq}
and~\eqref{eq:Dgen-v-uniq} to define the corresponding
\emph{generalised Ricci tensor} as 
\begin{equation}
\label{eq:GenRic+}
   \GenRic_{a\bar{b}} \,w_+^a 
      = \left[\Dgen_a , \Dgen_{\bar{b}} \right] w_+^a ,
\end{equation}
or as\footnote{Note that naively one might expect these definitions to
  give distinct tensors. However one can check that compatibility with
  the $O(p,q)\times O(q,p)$ structure means that the two agree.}
\begin{equation}
\label{eq:GenRic-}
   \GenRic_{\bar{a}b} \,w_-^{\bar{a}}
      = \left[\Dgen_{\bar{a}} , \Dgen_b \right] w_-^{\bar{a}} .
\end{equation}
Note that the index contractions are precisely what is needed to
guarantee uniqueness. 

It is not possible to contract the remaining two indices in the
generalised Ricci. Nonetheless, there does exist a notion of
generalised scalar curvature, but to define it we need the help of
spinors and the operators in~\eqref{eq:Dgen-spin-uniq}. We can obtain
the generalised Ricci again from either 
\begin{equation}
\label{eq:GenRicSpin}
\begin{aligned}
   \tfrac12 \GenRic_{a\bar{b}} \gamma^a \epsilon^+ 
      &= \left[\gamma^a \Dgen_a , \Dgen_{\bar{b}} \right] \epsilon^+ ,\\
   \tfrac12 \GenRic_{\bar{a}b} \gamma^{\bar{a}} \epsilon^- 
      &= \left[\gamma^{\bar{a}} \Dgen_{\bar{a}} , \Dgen_b \right] \epsilon^- .
\end{aligned}
\end{equation}
However, now we also find a \emph{generalised curvature scalar}
\begin{equation}
\label{eq:GenS+}
\begin{aligned}
	-\tfrac14 \GenS \epsilon^+ &= \big( \gamma^a \Dgen_a \gamma^b \Dgen_b 
	- \Dgen^{\bar{a}}\Dgen_{\bar{a}} \big) \epsilon^+ ,
\end{aligned}
\end{equation}
or alternatively,
\begin{equation}
\label{eq:GenS-}
\begin{aligned}
   -\tfrac14 \GenS \epsilon^- &= \big( 
     \gamma^{\bar{a}} \Dgen_{\bar{a}} \gamma^{\bar{b}} \Dgen_{\bar{b}} 
     - \Dgen^{a}\Dgen_{a} \big) \epsilon^- .
\end{aligned}
\end{equation}
Again, note the need to use the correct combinations of the operators
in these definitions so that all the undetermined components drop out.   

The fact that $\GenS$ is indeed a scalar and not itself an operator
might not be immediately apparent, so it is useful to work out the
explicit form of these curvatures. This can be done by again choosing
the two orthogonal frames to be aligned, $e_a^+=e_a^-$, to find 
\begin{equation}
\label{eq:GenRicFull}
	\GenRic_{ab} = \Ric_{ab} - 
	\tfrac{1}{4} H_{acd}H_b{}^{cd}+  2 \LC_a \LC_b \phi
	 + \tfrac{1}{2} \ee^{2\phi} \LC^c (\ee^{-2\phi} H_{cab}),
\end{equation}
and for the scalar
\begin{equation}
\label{eq:GenSFull}
\begin{aligned}
	\GenS = \Scalar +4 \LC^2 \phi -4 (\der \phi)^2  
	- \tfrac{1}{12} H^2 .
\end{aligned}
\end{equation}
From these expressions it is clear that we have obtained genuine
tensors which are uniquely determined by the torsion conditions, as
desired. Furthermore, comparing with~\cite{Siegel:1993xq,Siegel:1993th}
we see that locally these are the same tensors that appear in
Siegel's formulation. The expressions~\eqref{eq:GenRicFull}
and~\eqref{eq:GenSFull} also appear in the discussion of~\cite{JLP}.


\section{Type II supergravity as $O(9,1)\times O(1,9)$ generalised gravity}
\label{sec:IIasgen}

Having established the necessary elements of generalised geometry we
need, let us now show how the dynamics and supersymmetry
transformations of type II supergravity theories are encoded by an
$O(9,1)\times O(1,9)$ structure with a compatible, torsion-free
generalised connection. An outcome of this will be a formulation of
type II supergravity with manifest local $O(9,1)\times O(1,9)$
symmetry. 

In the following we will consider the full ten-dimensional supergraviy
theory so that the relevant generalised structure is
$O(10,10)\times\bbR^+$. However, one can equally well consider
compactifications of theory of the form $\bbR^{9-d,1}\times M$ 
\begin{equation}
   \dd s_{10}^2 = \dd s^2(\bbR^{9-d,1}) + \dd s_d^2 , 
\end{equation}
where $\dd s^2(\bbR^{9-d,1})$ is the flat metric on $\bbR^{9-d,1}$ and $\dd
s_d^2$ is a general metric on the $d$-dimensional manifold $M$. The
relevant structure is then the $O(d)\times O(d)\subset
O(d,d)\times\bbR^+$ generalised geometry on $M$. Below we will focus
on the $O(10,10)\times\bbR^+$ case. The compactification case
follows essentially identically. 


\subsection{NSNS and fermionic supergravity fields}
\label{sec:NSNS-fields}

From the discussion of section~\ref{sec:OdOd} we see that an
$O(9,1)\times O(1,9)\subset O(10,10)\times\bbR^+$ generalised
structure is parametrised by a metric $g$ of signature $(9,1)$, a
two-form $B$ patched as in~\eqref{eq:Bpatch} and a dilaton
$\phi$, that is, at each point $x\in M$
\begin{equation}
   \{ g, B, \phi \} 
      \in \frac{O(10,10)}{O(9,1)\times O(1,9)}\times\bbR^+ . 
\end{equation}
Thus it precisely captures the NSNS bosonic fields of type II
theories by packaging them into the generalised metric and conformal
factor $(G,\Phi)$. As in~\cite{GMPW}, the infinitesimal bosonic
symmetry transformation~\eqref{eq:NSsyms} is naturally encoded as the
Dorfman derivative by $V = v + \lambda$ 
\begin{equation}
	\delta_V G = \Lgen_V G,  \qquad \qquad \delta_V \Phi = \Lgen_V \Phi
\end{equation}
and the algebra of these transformations is given by the Courant bracket.

The type II fermionic degrees of freedom fall into spinor and
vector-spinor representations of
$\Spin(9,1)\times\Spin(1,9)$\footnote{Since the underlying manifold
  $M$ is assumed to possess a spin structure, we are free to promote
  $O(9,1) \times O(1,9)$ to $\Spin(9,1) \times\Spin(1,9)$. Here 
  will ignore more complicated extended spin structures that can arise in
  generalised geometry as described in~\cite{chris}.}. Let 
$S(C_+)$ and  $S(C_-)$ denote the $\Spin(9,1)$ spinor bundles
associated to the sub-bundles $C_\pm$ write $\gamma^a$ and
$\gamma^{\bar{a}}$ for the corresponding gamma matrices. Since we are
in ten dimensions, we can further decompose into spinor bundles
$S^\pm(C_+)$ and $S^\pm(C_-)$ of definite chirality under $\gamma^{(10)}$. 

The gravitino degrees of freedom then correspond to 
\begin{equation}
\label{eq:gravitino-rep}
\begin{aligned}
   \psi^+_{\bar{a}} &\in \Gs{C_-\otimes S^\mp(C_+)} , &&&&&
   \psi^-_a &\in \Gs{C_+\otimes S^+(C_-)} ,
\end{aligned}
\end{equation}
where the upper sign on the chirality refers to type IIA and the lower
to type IIB. Note that the vector and spinor parts of the gravitinos
transform under different $\Spin(9,1)$ groups. For the dilatino
degrees of freedom one has 
\begin{equation}
\label{eq:fermi-reps}
\begin{aligned}
   \rho^+ &\in \Gs{S^\pm(C_+)} , &&&&&
   \rho^- &\in \Gs{S^+(C_-)} , 
\end{aligned}
\end{equation}
where again the upper and lower signs refer to IIA and IIB
respectively. Similarly the supersymmetry parameters are sections 
\begin{equation}
\label{eq:fermi-reps}
\begin{aligned}
   \epsilon^+ &\in \Gs{S^\mp(C_+)} , &&&&&
   \epsilon^- &\in \Gs{S^+(C_-)} . 
\end{aligned}
\end{equation}
In terms of the string spectrum these gravitino and dilatino
representations just correspond to the explicit left- and right-moving
fermionic states of the superstring and, in a supergravity context
were discussed, for example, in~\cite{Hassan}.


\subsection{RR fields}

As is known from studying the action of T-duality, the RR field
strengths transform as $\Spin(10,10)$
spinors~\cite{Hassan,HT,BMZ,Fukuma:1999jt}. Here, the
patching~\eqref{eq:Apatch} of $A_{(i)}$ on $U_i\cap U_j$ implies that
the polyform $F_{(i)}=\dd A_{(i)}$ is patched as
in~\eqref{eq:Odd-spinor-patch}, and hence, as generalised spinors, 
\begin{equation}
   F\in \Gs{S^\pm_{(1/2)}} ,
\end{equation}
where the upper sign is for type IIA and the lower for type
IIB. Furthermore, we see that the RR field strengths $F^{(B)}_{(n)}$ 
that appear in the supergravity~\eqref{eq:RR} are simply $F$ expressed
in a split frame as in~\eqref{eq:PsiB}
\begin{equation}
   F^{(B)} = \ee^{B_{(i)}} \wedge F_{(i)} 
      = \ee^{B_{(i)}} \wedge \sum _n \dd A^{(n-1)}_{(i)} . 
\end{equation}
Note that the additional gauge transformations $\dd\hat{\Lambda}$
in~\eqref{eq:Apatch} imply that $A_{(i)}$ does not globally define a
section of $S^\pm_{(1/2)}$. ``Geometrising'' this additional gauge
symmetry is the subject of $E_{d(d)}$ generalised
geometry~\cite{CSW2}. Since $A_{(i)}$ is still locally a generalised spinor on the patch $U_i$ we can perform the same operations on it as we do on $F$ in the remainder of this subsection.

Given the generalised metric structure, we can also write $F$ in terms
of $\Spin(9,1)\times\Spin(1,9)$ representations. One has the
decomposition
$\Cliff(10,10;\bbR)\simeq\Cliff(9,1;\bbR)\otimes\Cliff(1,9;\bbR)$ with 
\begin{equation}
   \Gamma^A = \begin{cases}
      \gamma^a \otimes \id & \text{for $A = a$} \\
      \gamma^{(10)} \otimes \gamma^{\bar{a}} \gamma^{(10)} 
         & \text{for $A =\bar{a} + d$}
      \end{cases} . 
\end{equation}
and hence we can identify\footnote{In fact $S_{(p)}\simeq
  S(C_+)\otimes S(C_-)$ for any $p$, but here we focus on the case of
  interest $p =\tfrac12$} 
\begin{equation}
   S_{(1/2)}\simeq S(C_+)\otimes S(C_-) . 
\end{equation}
Using the spinor norm on $S(C_-)$ we can equally well view $F \in
\Gs{S_{(1/2)}}$ as a map from section of $S(C_-)$ to sections of $S(C_+)$. We
denote the image under this isomorphism as
\begin{equation}
\bisp{F} : S(C_-) \rightarrow S(C_+) . 
\end{equation}
We have that $F \in \Gs{S(C_+)\otimes S(C_-)}$ naturally has spin
indices $F^{\alpha \bar\alpha}$, while $\bisp{F}$ naturally has
indices $F^\alpha{}_{\bar\alpha}$. The isomorphism simply corresponds
to lowering an index with the $\Cliff(9,1;\bbR)$ intertwiner
$C_{\bar\alpha \bar\beta}$. The conjugate map, $\bisp{F}^T : S(C_+)
\rightarrow S(C_-)$, is given by 
\begin{equation}
   \bisp{F}^T = (C \bisp{F} C^{-1})^T ,
\end{equation}
which corresponds to lowering the other index on $F^{\alpha
  \bar\alpha}$ and taking the transpose. 

We now give the relations between the components of the
$\Spin(d,d)\times\mathbb{R}^+$ spinor in all relevant frames. Note
first that if the bases are aligned so that $e^+=e^-=e$ then the
$\Spin(9,1)\times\Spin(1,9)$ 
basis~\eqref{eq:OdOdexplicit} is a split conformal basis and we have a
$\Spin(9,1)\subset Spin(9,1)\times Spin(1,9)$ structure. We can then use the
isomorphism $\Cliff(9,1;\bbR)\simeq \Lambda^\bullet T^*M$ to write
$F^{(B,\phi)}$ as a spinor bilinear
\begin{equation}
   \slashed{F}^{(B,\phi)} 
      = \sum_n \tfrac{1}{n!} F^{(B,\phi)}_{a_1 \dots a_n} 
           \gamma^{a_1 \dots a_n} .
\end{equation}
More generally if the frames are related by Lorentz transformations
$e^\pm_a=\Lambda^{\pm b}_a e_a$ and we write $\Lambda^\pm$ for the
corresponding $\Spin(9,1§)$ transformations then we can define $\bisp{F}$ explicitly as
\begin{equation}
   \bisp{F}
      = \Lambda^+\slashed{F}^{(B,\phi)} (\Lambda^-)^{-1} ,
\end{equation}
which concretely realises the isomorphism between $F^{(B,\phi)}$ and $\bisp{F}$.

This map can easily be inverted and so we can write the components of
$F \in \Gs{S_{(1/2)}}$ in the coordinate frame as 
\begin{equation}
\begin{split}
   F_{(i)} &=  \ee^{-B_{(i)}} \wedge F^{(B)} 
      = \ee^{-\phi} \ee^{-B_{(i)}} \wedge F^{(B, \phi)} \\
      &= \ee^{-\phi} \ee^{-B_{(i)}} \wedge 
         \sum_n \Big[ \frac{1}{32 (n!)} (-)^{[n/2]}
		\tr \Big( \gamma_{(n)} (\Lambda^+)^{-1} 
                \bisp{F} \Lambda^-
		\Big) \Big] .
\end{split}
\end{equation}
This chain of equalities relates the components of $F$ in all the
frames we have discussed. 

Finally, we note that the self-duality conditions satisfied by the RR
field strengths $F\in \Gs{S^\pm_{(1/2)}}$ become a chirality condition
under the operator $\Gamma^{(-)}$ defined
in~\eqref{eq:OdxOd-chirality} 
\begin{equation}
	\Gamma^{(-)} F = - F ,
\end{equation}
as discussed in~\cite{Rocen:2010bk,Hohm:2011zr}. 


\subsection{Supersymmetry variations}
\label{sec:susy}

We now show that the supersymmetry variations can be written in a
simple, locally $\Spin(9,1)\times Spin(1,9)$ covariant form using the
torsion-free compatible connection $\Dgen$. 

We start with the fermionic variations~\eqref{eq:susy-fer}. Looking at
the expressions~\eqref{eq:Dgen-spin-uniq},  we see that the uniquely
determined spinor operators allow us to write the supersymmetry
variations compactly as 
\begin{equation}
\label{eq:Dgen-susy}
\begin{aligned}
   \delta \psi^+_{\bar{a}} &= \Dgen_{\bar{a}}\epsilon^+ + \tfrac{1}{16} 
\bisp{F}\gamma_{\bar{a}} \epsilon^-
           ,           \\
   \delta \psi^-_a &= \Dgen_a\epsilon^- 
   	+ \tfrac{1}{16} \bisp{F}^T \gamma_{a} \epsilon^+
           ,\\
   \delta\rho^+ 
      &= \gamma^a\Dgen_a\epsilon^+ 
    , \\
   \delta\rho^- 
      &= \gamma^{\bar{a}}\Dgen_{\bar{a}}\epsilon^- 
     , 
\end{aligned}
\end{equation}
where we have also used the results from the previous section to add
the RR field strengths to the gravitino variations.  

For the bosonic fields, we need the variation of a generic
$\Spin(9,1)\times\Spin(1,9)$ frame~\eqref{eq:OdOdexplicit}. Note that this
means defining the variation of a pair of orthonormal bases
$\{e^{+a}\}$ and $\{e^{-\bar{a}}\}$ whereas the conventional
supersymmetry variations~\eqref{eq:susy-bos} are given in terms of a
single basis $\{e^a\}$. The only possibility, compatible with the
$\Spin(9,1)\times\Spin(1,9)$ representations of the fermions, is to take  
\begin{equation}
\label{eq:Evar}
\begin{aligned}
   \tilde\delta\hat{E}^+_a 
      &= (\delta\log\Phi) \hat{E}^+_a 
         - (\delta\Lambda^+_{a\bar{b}}) \hat{E}^{-\bar{b}} , \\
   \tilde\delta\hat{E}^-_{\bar{a}} 
      &= (\delta\log\Phi) \hat{E}^-_{\bar{a}} 
         - (\delta\Lambda^-_{\bar{a}b}) \hat{E}^{+b}  , 
\end{aligned}
\end{equation}
where 
\begin{equation}
\begin{aligned}
   \delta\Lambda^+_{a\bar{a}}
      &= \bar{\epsilon}^+ \gamma_a \psi^+_{\bar{a}} 
           + \bar{\epsilon}^- \gamma_{\bar{a}} \psi^-_a , \\
   \delta\Lambda^-_{a\bar{a}}
      &= \bar{\epsilon}^+ \gamma_a \psi^+_{\bar{a}} 
           + \bar{\epsilon}^- \gamma_{\bar{a}} \psi^-_a , 
\end{aligned}
\end{equation}
and 
\begin{equation}
\label{eq:Phivar}
   \delta\log\Phi 
       = - 2\delta\phi + \tfrac{1}{2}\delta\log(-g) 
       = \bar{\epsilon}^+ \rho^+ + \bar{\epsilon}^- \rho^- . 
\end{equation}
Note that the variation of the basis~\eqref{eq:Evar} is
by construction orthogonal to the $\Spin(9,1)\times\Spin(1,9)$ action. This is
because it is impossible to construct an $\Spin(9,1)\times\Spin(1,9)$ tensor
linear in $\psi^+_{\bar{a}}$ and $\psi^-_a$ with two indices of the
same type, that is $L^+_{ab}$ or $L^-_{\bar{a}\bar{b}}$. 

The corresponding variations of the frames $\hat{e}^\pm$ are
\begin{equation}
\label{eq:doubled-variation}
\begin{aligned}
   \tilde\delta e_\mu^{+a} 
      &= \bar{\epsilon}^+ \gamma_\mu \psi^{+a} 
           + \bar{\epsilon}^- \gamma^a \psi^-_\mu , \\
   \tilde\delta e_\mu^{-\bar{a}} 
      &= \bar{\epsilon}^+ \gamma^{\bar{a}} \psi^+_\mu 
           + \bar{\epsilon}^- \gamma_\mu \psi^{-\bar{a}} , 
\end{aligned}
\end{equation}
which both give 
\begin{equation}
   \tilde\delta g_{\mu \nu} 
      = 2 \bar{\epsilon}^+ \gamma^{\phantom{+}}_{(\mu} \psi^+_{\nu)}
           + 2\bar{\epsilon}^- \gamma^{\phantom{+}}_{(\mu} \psi^-_{\nu)} , 
\end{equation}
as required, but, when setting the frames equal so $e^{+a}=e^a$ and
$e^{-\bar{a}}=e^{\bar{a}}$, differ by Lorentz transformations from the
standard form~\eqref{eq:susy-bos}
\begin{equation}
\label{eq:doubled-variation}
\begin{aligned}
   \tilde\delta e_\mu^{+a} 
      &= \delta e_\mu^{+a} - \big(
           \bar{\epsilon}^+ \gamma^a \psi^{+b}
           - \bar{\epsilon}^+ \gamma^b \psi^{+a} \big) e^+_{\mu b} , \\  
   \tilde\delta e_\mu^{-\bar{a}} 
      &= \delta e_\mu^{+\bar{a}} - \big(
           \bar{\epsilon}^- \gamma^{\bar{a}} \psi^{-\bar{b}}
           - \bar{\epsilon}^- \gamma^{\bar{b}} \psi^{-\bar{a}} 
           \big) e^-_{\mu \bar{b}} . 
\end{aligned}
\end{equation}
This can also be expressed in terms of the generalised metric $G_{AB}$ as
\begin{equation}
\label{eq:deltaG}
   \delta G_{a\bar{a}} = \delta G_{\bar{a} a} 
      = 2 \left( \bar{\epsilon}^+ \gamma_a \psi^+_{\bar{a}} 
         + \bar{\epsilon}^- \gamma_{\bar{a}} \psi^-_a \right) .
\end{equation}
The variation of the RR potential $A$ can be written as a bispinor
\begin{align}
   \tfrac{1}{16} (\delta \bisp{A}) 
      = \big( \gamma^a \epsilon^+ \bar\psi^-_a
         - \rho^+\bar\epsilon^- \big) 
         \mp \big( \psi^+_{\bar{a}}\bar\epsilon^- \gamma^{\bar{a}}
         + \epsilon^+\bar\rho^- \big) ,
\end{align}
where the upper sign is for type IIA and the lower for type IIB.


\subsection{Equations of motion}
\label{sec:gen-eom}

Finally, we rewrite the supergravity equations of
motion~\eqref{eq:bose-eom} and~\eqref{eq:fermi-eom} with local
$\Spin(9,1)\times\Spin(1,9)$ covariance, using the generalised notions
of curvature obtained in section~\ref{sec:curv}. 

From the generalised Ricci tensor \eqref{eq:GenRicFull}, we find that
the equations of motion for $g$ and $B$ can be written as 
\begin{equation}
\label{eq:G-eom}
   \GenRic_{a\bar{b}} 
      + \tfrac{1}{16}\Phi^{-1}\mukai{F}{\Gamma_{a \bar{b}} F} = 0,
\end{equation}
where we have made use of the Mukai pairing defined
in~\eqref{eq:mukai}\footnote{Note that $\mukai{F}{\Gamma_{a \bar{b}}
    F} \in \Gs{(\det{T^*M}) \otimes C_+ \otimes C_-}$ so $ \Phi^{-1}
  \mukai{F}{\Gamma_{a \bar{b}} F} \in \Gs{C_+ \otimes C_-}$} to
introduce the RR fields in a $\Spin(9,1)\times\Spin(1,9)$ covariant
manner. 

The equation of motion for $\phi$ does not involve the RR fields, so it is simply given by the generalised scalar curvature \eqref{eq:GenSFull}
\begin{equation}
\label{eq:Phi-eom}
   \GenS = 0 .
\end{equation}

Using definition~\eqref{eq:gen-dd} and equation~\eqref{eq:gen-dd-torsion} we can write the equation of motion for the RR fields in the familiar form
\begin{equation}
\label{eq:F-eom}
	\tfrac12 \Gamma^A \Dgen_A F = \dd F = 0 ,
\end{equation}
where the first equality serves as a reminder that this definition of
the exterior derivative is fully covariant under
$\Spin(d,d)\times\mathbb{R}^+$.  

We also have the bosonic pseudo-action~\eqref{eq:NSaction} which takes
the simple form\footnote{Up to integration by parts of the $\nabla^2
  \phi$ term} 
\begin{equation}
\label{eq:SB-gen}
\begin{aligned}
   S_B = \frac{1}{2\kappa^2}\int \left( \,  \Phi \, \GenS 
            + \tfrac{1}{4} \mukai{F}{\Gamma^{(-)} F} \right) ,
\end{aligned}
\end{equation}
using the density $\Phi$. Note that the Mukai pairing is a top-form
which can be directly integrated. 

The fermionic action~\eqref{eq:fermi-action} is given by
\begin{equation}
\begin{aligned}
   S_F = -\frac{1}{2\kappa^2}\int 2 \Phi \Big[ 
         \bar\psi^{+\bar{a}} &\gamma^b \Dgen_b \psi^+_{\bar{a}}
         + \bar\psi^{-a} \gamma^{\bar{b}} \Dgen_{\bar{b}} \psi^-_{a}  \\
         &+ 2 \bar\rho^+ \Dgen_{\bar{a}} \psi^{+\bar{a}}
         + 2 \bar\rho^- \Dgen_a \psi^{-a} \\
         &- \bar\rho^+ \gamma^a \Dgen_a \rho^+ 
         - \bar\rho^- \gamma^{\bar{a}} \Dgen_{\bar{a}} \rho^-  \\
         &-\tfrac{1}{8} \Big( \bar\rho^+ \bisp{F}\rho^-
         + \bar\psi^+_{\bar{a}} \gamma^a \bisp{F}\gamma^{\bar{a}} \psi^-_a \Big)
         \Big].
\end{aligned}
\end{equation}
Varying this with respect to the fermionic fields leads to the
generalised geometry version of \eqref{eq:fermi-eom} 
\begin{equation}
\begin{aligned}
   \gamma^b \Dgen_b \psi^+_{\bar{a}} - \Dgen_{\bar{a}} \rho^+ 
      &= + \tfrac{1}{16} \gamma^b \bisp{F} \gamma_{\bar{a}} \psi^-_b  ,\\
   \gamma^{\bar{b}} \Dgen_{\bar{b}} \psi^-_{a} - \Dgen_{a} \rho^- 
      &= + \tfrac{1}{16} 
         \gamma^{\bar{b}} \bisp{F}^T \gamma_{a} \psi^+_{\bar{b}}  ,\\
    \gamma^a \Dgen_a \rho^+ - \Dgen^{\bar{a}} \psi^+_{\bar{a}}
       &= - \tfrac{1}{16} \bisp{F} \rho^-  ,\\
    \gamma^{\bar{a}} \Dgen_{\bar{a}} \rho^- - \Dgen^a \psi^-_a
       &= - \tfrac{1}{16} \bisp{F}^T \rho^+ ,
\end{aligned}
\end{equation}
and it is straightforward to verify that by applying a supersymmetry
variation~\eqref{eq:Dgen-susy} we recover the bosonic equations of
motion~{\eqref{eq:G-eom}-\eqref{eq:F-eom}}. 

We have thus rewritten all the supergravity equations from
section~\ref{sec:eom} in terms of torsion free generalised connections
and therefore as manifestly covariant under local
$\Spin(9,1)\times\Spin(1,9)$ transformations.


\section{Conclusions and discussion}
\label{sec:conc}


Starting with a generalised tangent space with a
$O(10,10)\times\bbR^+$ structure, we have shown that type II
supergravity can be understood as a gravitational theory for a
$\Spin(9,1)\times\Spin(1,9)$ substructure. It is defined using a
torsion-free compatible connection $\Dgen$, in 
direct analogy to conventional gravity with a Levi--Civita
connection. Our reformulation includes the leading fermionic equations
of motion and action, and all the supersymmetry variations. The theory
has a local $\Spin(9,1)\times\Spin(1,9)$ covariance together with an
extension of the diffeomorphism group by the $B$-field gauge
transformations. 

As we mentioned in the introduction,  both in Siegel's
formulation~\cite{Siegel:1993xq,Siegel:1993th} and in double field
theory~\cite{Hull:2009mi}, if one requires the action to be gauge
invariant, one imposes the condition that on each coordinate patch the
fields are independent of half the doubled coordinates. Thus locally,
the generalised geometry and double field theory descriptions are
completely equivalent, and our reformulation also gives the fermionic
equations of motion, action and the supersymmetry variations in
double field theory. 

The relation of our formalism to Siegel's is interesting since naively
he uses a $\GL(d,\bbR)\times\GL(d,\bbR)$ structure rather than a
$O(p,q)\times O(q,p)$ structure. However, his
$\GL(d,\bbR)\times\GL(d,\bbR)$ connection is also required preserve 
the $O(d,d)$ metric and the volume measure ($\Phi$ in our
notation). Thus in fact, the connection is compatible with the common
subgroup of $O(d,d)\times\bbR^+$ and the appropriate embedding of
$\GL(d,\bbR)\times\GL(d,\bbR)$ in $\GL(2d,\bbR)$, namely $O(p,q)\times
O(q,p)$. This explains the agreement of our curvature
tensors. Similarly the lack of covariance of Siegel's (modified)
putative Riemann tensor is a reflection of the non-tensorial nature
described in~\eqref{eq:Dgen-comm-linear}.

One of the most remarkable properties of the reformulation is that
supersymmetry was not used in the construction of the connection $D$
and yet it has precisely the properties necessary for the
supersymmetry algebra to close. For instance, from~\eqref{eq:deltaG}
and \eqref{eq:Dgen-susy} we see that the double variation of
generalised metric is just a Dorfman derivative $\Lgen_V G$, that is
simply a diffeomorphism plus gauge transformation, precisely because 
$\Dgen$ is torsion-free. This is strongly suggestive of the fact that
the construction has a natural supersymmetric extension. 

As will be reported in~\cite{CSW2}, similar generalised geometrical
constructions, using $E_{d(d)}\times\bbR^+$ structures, describe
eleven-dimensional supergravity restricted to $d$-dimensional 
spacetimes. In fact, there is evidence that several pure supergravity
theories, in varying dimensions, can be formulated this
way. This leads to the question of why there is such a general
relationship between supergravity and versions of generalised geometry
and which types of structure groups can appear.  

There are a number of other directions for which this formulation may
prove useful. One is the description of higher-derivative correction
terms to the theory, assuming the generalised structure is not
broken. Another is the explicit construction of supergravity
backgrounds, for instance as spaces with particular special structures
on the generalised tangent space. One can also connect this work to
that on non-geometrical backgrounds. For example, in~\cite{ALLP} the
NSNS action is rewritten in terms of a ``non-geometrical'' flux
$Q$. In the formalism of this paper, this amounts to evaluating
$\GenS$ in a different frame from the standard split frame used
in~\eqref{eq:GenSFull}. One takes instead $\hat{E}_a =
\ee^{-2\phi}(\det e) \hat{e}_a$ and $E^a=\ee^{-2\phi}(\det e)\left(e^a
   + \beta \hat{e}^a \right)$ where $\beta$ is a bivector
$\beta\in\Gs{\Lambda^2TM}$. Locally, from a generalised geometrical
perspective, these are equivalent. However, given the
patching~\eqref{eq:Odd-patch}, the new frame is not, generically,
globally defined in a conventional generalised geometry. The
suggestion though is that on a non-geometrical background (patched for
instance by a T-duality) it may be possible to make some global notion
of such a frame. 

Perhaps, indeed, the most interesting question is whether there
can be any extension of the generalised geometrical picture described 
here relevant to such exotic string backgrounds. One that moves away
from a conventional ten-dimensional manifold, while still retaining
``geometrical'' notions of, for example, connections and curvatures,
as some sensible limit of the full string theory.


\acknowledgments

We would like to thank Mariana Gra\~{n}a and Chris Hull for helpful
discussions. C.~S-C.~is supported by an STFC PhD studentship. A.~C.~is
supported by the Portuguese Funda\c c\~ao para a Ci\^encia e a
Tecnologia under grant SFRH/BD/43249/2008. D.~W.~also thanks
CEA/Saclay and the Mitchell Institute for 
Fundamental Physics and Astronomy at Texas A\&M for hospitality during the
completion of this work. 


\appendix


\section{Supergravity Conventions}
\label{app:conv}

Our conventions largely follow \cite{democratic} but we include a list for completeness. The only difference which is not purely notational is that we take the opposite sign for the Riemann tensor, as discussed in appendix~\ref{app:LC}. The metric has the mostly plus signature $(-++ \dots +)$. We use the indices $\mu,\nu,\lambda \dots $ as the spacetime coordinate indices and $a,b,c \dots$ for the tangent space indices. We take symmetrisation of indices with weight one. Our conventions for forms are
\begin{align}
	\omega_{(k)} &= \tfrac{1}{k!} \omega_{\mu_1 \dots \mu_k} 
		\dd x^{\mu_1} \wedge \dots \wedge \dd x^{\mu_k} \nonumber ,\\
	\omega_{(k)} \wedge \eta_{(l)} &= \tfrac{1}{(k+l)!} \left( \tfrac{(k+l)!}{k! \, l!}
		 \omega_{[\mu_1 \dots \mu_k} \eta_{\mu_{k+1} \dots \mu_{k+l}]} \right)
		\dd x^{\mu_1} \wedge \dots \wedge \dd x^{\mu_{k+l}} \nonumber ,\\
	* \omega_{(k)} &= \tfrac{1}{(10-k)!} 
		\left( \tfrac{1}{k!} \sqrt{-g} \epsilon_{\mu_1 \dots \mu_{10-k} \nu_1 \dots \nu_k} 
			\omega^{\nu_1 \dots \nu_k} \right)
			\dd x^{\mu_1} \wedge \dots \wedge \dd x^{\mu_{10-k}} ,
\end{align}
where $\epsilon_{01 \dots 9} = -\epsilon^{01 \dots 9} = +1$.
The gamma matrices have
\begin{align}
	\left\{ \gamma^\mu , \gamma^\nu \right\} = 2 g^{\mu\nu}, 
		& & \gamma^{\mu_1 \dots \mu_k} = \gamma^{[\mu_1} \dots \gamma^{\mu_k]} ,
\end{align}
and we use the anti-symmetric transpose intertwiner
\begin{align}
	C \gamma^\mu C^{-1} = - (\gamma^\mu)^T , & & C^T = - C ,
\end{align}
to define the Majorana conjugate as $\bar\epsilon = \epsilon^T C$. This leads to the formulae
\begin{align}
	C \gamma^{\mu_1 \dots \mu_k} C^{-1} = (-)^{[(k+1)/2] } (\gamma^{\mu_1 \dots \mu_k})^T \nonumber ,\\ 
	\bar\epsilon \gamma^{\mu_1 \dots \mu_k} \chi 
        = (-)^{[(k+1)/2]} \bar\chi \gamma^{\mu_1 \dots \mu_k} \epsilon ,
\end{align}
where in the second equation the spinors $\epsilon$ and $\chi$ are anti-commuting. The top gamma is defined as 
\begin{equation}
	\gamma^{(10)} = \gamma^0 \gamma^1 \dots \gamma^9 = \tfrac{1}{10!} \epsilon_{\mu_1 \dots \mu_{10}} \gamma^{\mu_1 \dots \mu_{10}} ,
\end{equation}
and this gives rise to the equation
\begin{equation}
	\gamma_{\mu_1 \dots \mu_k} \gamma^{(10)} = (-)^{[k/2]} \tfrac{1}{(10-k)!} \sqrt{-g} 
		\epsilon_{\mu_1 \dots \mu_k \nu_1 \dots \nu_{10-k}} \gamma^{\nu_1 \dots \nu_{10-k}} ,
\end{equation}
which is also commonly written as
\begin{equation}
	\gamma^{(k)} \gamma^{(10)} = (-)^{[k/2]} * \gamma^{(10-k)} .
\end{equation}
We use Dirac slash notation with weight one so that for $\Psi \in \Gs{\Lambda^\bullet T^*M}$
\begin{equation}
	\slashed{\Psi} = \sum_k \tfrac{1}{k!} \Psi_{\mu_1 \dots \mu_k} \gamma^{\mu_1 \dots \mu_k} .
\end{equation}


\section{Metric structures, torsion and the Levi--Civita connection}
\label{app:LC}

In this appendix we briefly review the basic geometry that goes into
the construction of the Levi--Civita connection, as context for the
corresponding generalised geometrical analogues.  

Let $M$ be a $d$-dimensional manifold. We write $\{\hat{e}_a\}$ for a
basis of the tangent space $T_xM$ at $x\in M$ and $\{e^a\}$ be the
dual basis of $T^*_xM$ satisfying $i_{\hat{e}_a}e^b=\delta_a{}^b$. 
Recall that the frame bundle $F$ is the bundle of all bases
$\{\hat{e}^a\}$ over $M$, 
\begin{equation}
\label{eq:Fdef}
   F = \left\{ (x,\{\hat{e}_a\}) : \ \text{$x\in M$ and
        $\{\hat{e}_a\}$ is a basis for $T_xM$} \right\} .
\end{equation}
On each fibre of $F$ there is an action of $A^a{}_b\in\GL(d,\bbR)$,
given $v=v^a\hat{e}_a\in\Gs{T_xM}$, 
\begin{equation}
   v^a \mapsto v^{\prime a} = A^a{}_b v^b , \qquad
   \hat{e}_a\mapsto \hat{e}'_a=\hat{e}_b(A^{-1})^b{}_a . 
\end{equation}
giving $F$ the structure of a $\GL(d,\bbR)$ principal bundle.

The Lie derivative $\mathcal{L}_v$ encodes the effect of an
infinitesimal diffeomorphism. On a vector field $w$ it is equal to the
Lie bracket 
\begin{equation}
\label{eq:LDdef}
   \mathcal{L}_v w=-\mathcal{L}_w v=\BLie{v}{w} , 
\end{equation}
while on a general tensor field $\alpha$ one has, in coordinate
indices, 
\begin{equation}
\label{eq:LDgen}
\begin{aligned}
   \mathcal{L}_v\alpha^{\mu_1\dots \mu_p}_{\nu_1\dots \nu_q} 
      &= v^\mu \der_\mu \alpha^{\mu_1\dots \mu_p}_{\nu_1\dots \nu_q} 
          \\ & \qquad
          + \left(\der_\mu v^{\mu_1}\right) 
               \alpha^{\mu \mu_2\dots \mu_p}_{\nu_1\dots \nu_q} 
          + \dots
          + \left(\der_\mu v^{\mu_p}\right) 
               \alpha^{\mu_1\dots \mu_{p-1} \mu}_{\nu_1\dots \nu_q}
          \\ & \qquad
          - \left(\der_{\nu_1} v^\mu\right) 
               \alpha^{\mu_1\dots \mu_p}_{\mu \nu_2\dots \nu_q} 
          - \dots
          - \left(\der_{\nu_q} v^\mu\right) 
               \alpha^{\mu_1\dots \mu_p}_{\nu_1\dots \nu_{q-1} \mu} .
\end{aligned} 
\end{equation}
Note that the terms on the second and third lines can be viewed as the
adjoint action of the $\mathfrak{gl}(d,\bbR)$ matrix $a^\mu{}_\nu=\der_\nu
v^\mu$ on the particular tensor field $\alpha$. This form will have an
analogous expression when we come to generalised geometry.  

Let $\nabla_\mu v^\nu=\der_\mu v^\nu+\omega_\mu{}^\nu{}_\lambda v^\lambda$ be a general
connection on $TM$. The torsion $T\in\Gs{TM\otimes\Lambda^2T^*M}$ of
$\nabla$ is defined by 
\begin{equation}
\label{eq:Tdef}
   T(v,w) = \nabla_v w - \nabla_w v - \BLie{v}{w} . 
\end{equation}
or concretely, in coordinate indices, 
\begin{equation}
\label{eq:Tcomp}
   T^\mu{}_{\nu\lambda} = \omega_\nu{}^\mu{}_\lambda - \omega_\lambda{}^\mu{}_\nu , 
\end{equation}
while, in a general basis where $\nabla_\mu
v^a=\der_\mu v^a+\omega_\mu{}^a{}_bv^b$, one has  
\begin{equation}
\label{eq:Tcomp}
   T^a{}_{bc} = \omega_b{}^a{}_c - \omega_c{}^a{}_b
       + \BLie{\hat{e}_b}{\hat{e}_c}^a . 
\end{equation}
Since again it has a natural generalised geometric analogue, it is
useful to equivalently define the torsion in terms of the Lie
derivative. If $\mathcal{L}^\nabla_v\alpha$ is the analogue of the Lie
derivative~\eqref{eq:LDgen} but with $\der$ replaced by $\nabla$, and $(i_vT)^\mu{}_\nu=v^\lambda T^\mu{}_{\lambda\nu}$ then 
\begin{equation}
\label{eq:Tdef2}
   (i_vT)\alpha 
       = \mathcal{L}^\nabla_v\alpha - \mathcal{L}_v\alpha ,
\end{equation}
where we view $i_vT$ as a section of the $\mathfrak{gl}(d,\bbR)$
adjoint bundle acting on the given tensor field $\alpha$. 

The curvature of a connection $\nabla$ is given by the Riemann tensor $\Riem \in \Gs{\Lambda^2T^*M \otimes TM \otimes T^*M}$, defined by
\begin{equation}
\label{eq:Rdef}
\begin{aligned}
	\Riem(u,v)w &= [ \LC_u, \LC_v ] w - \LC_{[u,v]} w ,\\
	\Riem_{\mu\nu\phantom{\lambda}\rho}^{\phantom{\mu\nu}\lambda}v^{\rho} &= [ \nabla_\mu , \nabla_\nu ] v^\lambda - T^\rho{}_{\mu\nu}\nabla_{\rho}v^\lambda .
\end{aligned}
\end{equation}
The Ricci tensor
is the trace of the Riemann curvature
\begin{align}
\label{eq:Ricdef}
\Ric_{\mu\nu}=\Riem_{\lambda\mu\phantom{\lambda}\nu}^{\phantom{\mu\nu}\lambda} .
\end{align}
If the manifold admits a metric $g$ then the Ricci scalar is defined by
\begin{align}
\label{eq:Sdef}
\Scalar=g^{\mu\nu}\Ric_{\mu\nu} .
\end{align}

A $G$-structure is a principal sub-bundle $P\subset F$ with
fibre $G$. In the case of the metric $g$, the $G=O(d)$ sub-bundle is
formed by the set of orthonormal bases 
\begin{equation}
   \label{eq:Od-struc}
   P = \left\{ (x,\{\hat{e}_a\}) \in F :
        g(\hat{e}_a,\hat{e}_b)=\delta_{ab} \right\} ,
\end{equation}
related by an $O(d)\subset\GL(d,\bbR)$ action. (A Lorentzian  defines
a $O(d-1,1)$-structure and $\delta_{ab}$ is replaced by $\eta_{ab}$.)
At each point $x\in M$, the metric defines a point in the coset space 
\begin{equation}
   g|_x\in \GL(d,\bbR)/O(d) .
\end{equation}
In general the existence of a $G$-structure can impose topological
conditions on the manifold, since it implies that the tangent space
can be patched using only $G\subset\GL(d,\bbR)$ transition
functions. (For example, for even $d$, if $G=\GL(d/2,\bbC)$, the
manifold must admit an almost complex structure, while for
$G=\SL(d,\bbR)$ it must be orientable.) However, for $O(d)$ there is
no such restriction.  

A connection $\nabla$ is compatible with a $G$-structure $P\subset E$
if the corresponding connection of the principal bundle $E$ reduces to
a connection on $P$. This means that, given a basis $\{\hat{e}_a\}$,
one has a set of connection one-forms $\omega^a{}_b$ taking values 
in the adjoint representation of $G$ given by 
\begin{equation}
 \nabla_{\der/\der x^\mu} \hat{e}_a = \omega_\mu{}^b{}_a \hat{e}_b . 
\end{equation}
For a metric structure this is equivalent to the condition $\nabla
g=0$. If there exists a torsion-free compatible connection, the 
$G$-structure is said to be torsion-free or equivalently
integrable (to first order). In general this can further restrict
the structure, for instance in the case of $\GL(d/2,\bbC)$ it is
equivalent to the existence of a complex structure (satisfying the
Nijenhuis condition). However, for a metric structure no further
conditions are implied, and furthermore the torsion-free, compatible
connection, namely the Levi--Civita connection, is unique.



\end{document}